\documentclass[aps,prl,reprint,superscriptaddress,longbibliography,nofootinbib,floatfix]{revtex4-2}

\usepackage[utf8]{inputenc}
\usepackage[T1]{fontenc}
\usepackage{amsmath,amssymb,mathtools}
\usepackage{graphicx}
\usepackage{bbm}
\usepackage{pdfpages}
\usepackage{pgffor}

\makeatletter
\AtBeginDocument{\let\LS@rot\@undefined}
\makeatother

\usepackage[colorlinks=true,linkcolor=blue,citecolor=blue,urlcolor=blue]{hyperref}
\usepackage[capitalize]{cleveref}

\newcommand{\kb}{k_{\rm B}}
\newcommand{\upd}{\mathrm{d}}
\newcommand{\order}{\mathcal{O}}
\newcommand{\complexi}{\mathbbm{i}}
\renewcommand{\vec}[1]{\mathbf{#1}}
\renewcommand{\Re}{\operatorname{Re}}
\renewcommand{\Im}{\operatorname{Im}}

\begin{document}

\title{Acoustic dip in the effective temperature of hot Brownian motion}

\author{Mayank Srivastava}
\email{mayanksrivastava94@gmail.com}
\affiliation{Indian Institute of Science Education and Research Mohali,
Knowledge City, Sector 81, S. A. S. Nagar, Manauli-140306, India}

\author{Dipanjan Chakraborty}
\email{chakraborty@iisermohali.ac.in}
\affiliation{Indian Institute of Science Education and Research Mohali,
Knowledge City, Sector 81, S. A. S. Nagar, Manauli-140306, India}

\begin{abstract}
  A hot Brownian particle is driven by thermal fluctuations from a
  nonuniformly heated solvent and therefore obeys a
  fluctuation-dissipation relation with a frequency-dependent
  effective temperature. Existing theories assume an incompressible
  solvent. We show that finite sound speed qualitatively changes this
  effective temperature in the kinetic regime. Solving the
  compressible fluctuating-hydrodynamic problem for a heated sphere,
  we find that the effective noise temperature develops a pronounced
  acoustic dip when the sound wavelength becomes comparable to the
  particle radius. The dip occurs because weakly attenuated
  longitudinal modes carry mechanical energy away from the heated
  surface before it is dissipated, so that the dissipation is weighted
  by colder regions of the temperature field. Its position is set by
  $a\omega/c=\order(1)$, while its depth is controlled by the
  dimensionless viscous-acoustic attenuation length $a_c/a$. The
  result identifies an acoustic window in hot Brownian motion in which
  compressibility lowers, rather than raises, the kinetic noise
  temperature.
\end{abstract}

\maketitle

Hot Brownian motion\cite{Rings:2010iy,Radunz:2009ka,Chakraborty:2011kk,Rings:2011gj} describes the Brownian dynamics of a particle
maintained at an elevated a temperature from that of the surrounding
solvent. The solvent is then out of global thermal equilibrium, but
local equilibrium allows one to assign a space-dependent noise
strength to the fluctuating stress. The Brownian particle does not
sample this temperature field uniformly; it samples it through the
hydrodynamic dissipation associated with the translational and
rotational degress of freedom. This gives
rise to a frequnecy dependent effective temperatures for hot Brownian motion.
\cite{Falasco:2014iq,Srivastava:2018} Previous
calculations of these effective temperatures have treated the solvent
as incompressible. This is natural for the diffusive, long-time
Brownian regime, but it hides the role of sound modes in the high frequency
regime.

In this Letter we show that compressibility produces a qualitatively
new feature: an acoustic dip in the frequency-dependent effective
temperature of the thermal force acting on the particle. The dip
appears when the acoustic wavelength becomes comparable to the
particle size, while the acoustic attenuation length is still much
larger than the particle radius. In that regime longitudinal sound
modes are efficiently excited, but dissipate their energy away from
the hot particle. Since the effective temperature is a
dissipation-weighted average of the solvent temperature, this remote
dissipation lowers the noise temperature.

We consider a sphere of radius $a$ suspended in a solvent at ambient
temperature $T_0$. The particle is maintained at $T_0+\Delta T$, with
$\Delta T/T_0\ll1$, so that the steady temperature field is
\begin{equation}
T(r)=T_0+\Delta T\frac{a}{r} .
\label{eq:temperature_profile}
\end{equation}
We focus on the translational motion of the particle and assume no
slip at the particle surface. The fluid is described by linearized
fluctuating hydrodynamics \cite{Chow1972,Hauge1973,Velarde1974a}. The
deterministic stress is
\begin{equation}
\sigma_{ij}=-p\delta_{ij}+2\eta e_{ij}+\left(\mu-\frac{2}{3}\eta\right)\delta_{ij}\partial_k v_k ,
\end{equation}
where $\eta$ is the shear viscosity and $\mu$ is the bulk
viscosity. The fluctuating stress satisfies a local
fluctuation--dissipation relation
\begin{equation}
  \left\langle \tau_{ij}(\vec{r},t)\tau_{kl}(\vec{r'},t')
  \right\rangle=2 k_B T(\vec{r},t)
   C_{ijkl}\delta(\vec{r}-\vec{r}')\delta(t-t'),
\end{equation}
where
$C_{ijkl}=\eta(\delta_{ik}\delta_{jl}+\delta_{il}\delta_{jk})+\left(\mu-\frac{2}{3}\eta\right)
\delta_{ij}\delta_{kl}$. We assume a stationary temperature profile,
constant material coefficients evaluated at $T_0$, a no-slip boundary
condition, and the validity of the local equilibrium and continuum
hydrodynamics over the frequency range considered.

The particle obeys a generalized Langevin equation,
\begin{equation}
M\dot{\vec V}(t)=-\int_{-\infty}^{t}\zeta(t-t')\vec V(t')\,\upd t' +\vec \xi(t)+\vec F_{\rm ext}(t).
\end{equation}
The frequency-dependent noise temperature $\mathcal T(\omega)$ is defined by
\begin{equation}
\left\langle \xi_i(\omega)\xi_j(\omega')\right\rangle
=2\kb\mathcal T(\omega)\,\Re\zeta(\omega)\,
\delta_{ij}\delta(\omega+\omega') .
\end{equation}
For a nonisothermal fluid, fluctuating hydrodynamics gives\cite{Falasco:2014iq,Srivastava:2018}
\begin{equation}
\mathcal T(\omega)=
\frac{\int_a^\infty T(r)\phi(\vec{r},\omega)\,\upd \vec{r}}
{\int_a^\infty \phi(\vec{r},\omega)\,\upd \vec{r}},
\label{eq:effective_T_def}
\end{equation}
where $\phi(\vec{r},\omega)$ is the local rate viscous and compressional
dissipation generated by a unit-amplitude oscillatory velocity of the
particle. We therefore write
\begin{equation}
\Theta(\omega)\equiv\frac{\mathcal T(\omega)-T_0}{\Delta T}
=\frac{\mathcal N(\omega)}{\mathcal D(\omega)} .
\label{eq:Theta_def}
\end{equation}
Thus the problem reduces to finding where the hydrodynamic motion dissipates energy.

Compressibility introduces two hydrodynamic wavenumbers. The
transverse mode is controlled by
\begin{equation}
\alpha^2=\complexi\omega\rho/\eta,
\qquad
\alpha a=(1+\complexi)x,
\qquad
x=a\sqrt{\frac{\omega}{2\nu}},
\end{equation}
where $\nu=\eta/\rho$. The longitudinal mode is controlled by
\begin{equation}
\beta^2=\frac{\omega^2}{c^2-\complexi\omega(\mu+4\eta/3)/\rho},
\qquad
\beta a=u+\complexi v .
\end{equation}
Here $u$ fixes the acoustic wavelength and $v$ fixes the acoustic
attenuation length. It is useful to introduce
\begin{equation}
\begin{gathered}
y=ak_c=\frac{a\omega}{c},\qquad
\chi=\frac{\nu}{ac},\qquad
\ell=\frac{a_c}{a},\\
a_c=\frac{\mu+4\eta/3}{\rho c}.
\end{gathered}
\label{eq:dimensionless_variables}
\end{equation}
Then $x=(y/2\chi)^{1/2}$. The parameter $a_c$ is a viscous-acoustic
material length scale; the actual attenuation length is frequency
dependent and is set by $(\Im \beta)^{-1}$. The limit $y=\order(1)$ is
the acoustic crossover in which the sound wavelength is comparable to
the particle diameter up to a factor of order unity. The complete
problem is now described in terms of three dimensionless parameters:
$y,\chi$ and $\ell$, i.e the numerator and the denominator in
\cref{eq:effective_T_def} and therefore $\Theta(\omega) $ are
functions of only these three parameters.

The velocity field for an oscillating sphere in a compressible fluid
is a superposition of the transverse and longitudinal modes of Chow
and Hermans \cite{Chow1973a}. Carrying out the spatial integrations in
Eq.~\eqref{eq:effective_T_def} gives
\begin{equation}
|\Delta|^2\mathcal D(\omega)=\frac{8\pi}{3}a\eta V^2\,\overline{\mathcal D}(\omega),
\qquad
|\Delta|^2\mathcal N(\omega)=\frac{8\pi}{3}a\eta V^2\,\overline{\mathcal N}(\omega),
\end{equation}
where the denominator is
\begin{equation}
\overline{\mathcal D}=9\frac{1}{4\mathcal{R}_\eta}|F_\alpha|^2
\frac{(u^2+v^2)^2(u^2+v^2+2v)}{v}
+36x^4(1+x)|F_\beta|^2 .
\label{eq:Dbar}
\end{equation}
Here
\begin{equation}
F_\alpha=3-3\complexi a\alpha-a^2\alpha^2,
\qquad
F_\beta=3-3\complexi a\beta-a^2\beta^2 .
\end{equation}
The numerator can be decomposed as
\begin{equation}
\overline{\mathcal N}=I_1+I_2+I_3+2\Re I_{\alpha\beta},
\label{eq:N_decomposition}
\end{equation}
where $I_1$ is the transverse self-contribution, $I_2+I_3$ is the
longitudinal self-contribution including the divergence correction,
and $I_{\alpha\beta}$ is the transverse-longitudinal cross
term. Explicit forms are given in the Supplemental Material. This
decomposition will be used below to identify the origin of the
acoustic dip.

\begin{figure}[t]
  \includegraphics[width=\linewidth]{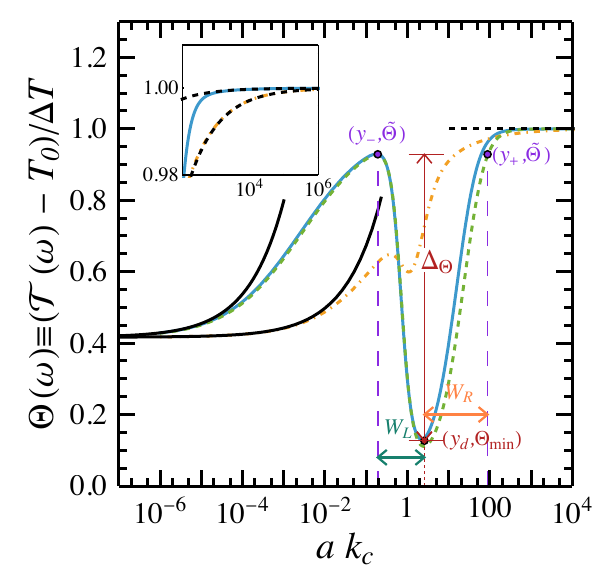}
  \caption{Frequency-dependent effective temperature
    $\Theta(\omega)=[\mathcal T(\omega)-T_0]/\Delta T$ plotted as a
    function of the dimensionless variable $y=ak_c=a\omega/c$. The
    curves correspond to a small particle with $a=5\sigma$ immersed in
    a Lennard--Jones solvent (dash-dotted line), a much larger
    particle $a=1000 \sigma$ in the same solvent, and for a particle
    of radius $1\,\mu{\rm m}$ in water. Compressibility produces a pronounced dip near
    $y=\order(1)$ when the acoustic wavelength is comparable to the
    particle size. The low-frequency limit is the incompressible
    hot-Brownian value $5/12$, while the high-frequency limit
    approaches the surface temperature, $\Theta\to1$.}
\label{fig:theta}
\end{figure}

~\cref{fig:theta} shows the central result. In the incompressible
theory the effective temperature crosses over monotnically from the hot-Brownian
value $\Theta(0)=5/12$ to the surface value $\Theta(\infty)=1$. For a
compressible fluid this crossover is nonmonotonic: a pronounced dip
appears near $a\omega/c=\order(1)$. For the Lennard-Jones fluid we use
$\rho=0.82$, $\eta=2.7$, $c=5$, and $a_c=1.07$ in reduced units. For
$a=5$ the dip is weak, while for $a=1000$ the reduced attenuation
length $\ell=a_c/a$ is of order $10^{-3}$ and the dip becomes deep. A
similar dip is obtained for a $1\,\mu{\rm m}$ particle in water using
$\rho=997\,{\rm kg/m^3}$, $\eta=0.89\times10^{-3}\,{\rm Pa\,s}$,
$\mu=2.5\times10^{-3}\,{\rm Pa\,s}$, and $c=1497\,{\rm m/s}$.

The limiting cases are useful for comparison. When $y\ll1$ the
particle is acoustically small. The longitudinal mode is weakly
excited, and the response reduces to the incompressible Stokes-flow
result. The first correction is the known incompressible one
\cite{Srivastava:2018},
\begin{equation}
\Theta(\omega)=\frac{5}{12}\left(1+a\sqrt{\frac{\omega\rho}{2\eta}}\right)+\cdots .
\end{equation}
At very high frequency the dissipation is localized close to the
particle surface, and therefore
\begin{equation}
\Theta(\omega)\to1 .
\end{equation}
In the true high-frequency compressible limit, both transverse and
longitudinal wavenumbers are large. If $s=[\eta/(\mu+4\eta/3)]^{1/2}=\sqrt{\mathcal{R}_\eta}$,
then $a\beta_1\simeq a\beta_2\simeq sx$ and
\begin{equation}
\Theta(\omega)=1-\frac{2s^2+1}{2s(2s+1)}\frac{1}{x}+\order(x^{-2}).
\label{eq:high_freq}
\end{equation}
The dip in Fig.~\ref{fig:theta} is therefore not a limiting asymptote;
it is an intermediate acoustic regime.

\begin{figure}[t]
\includegraphics[width=\linewidth]{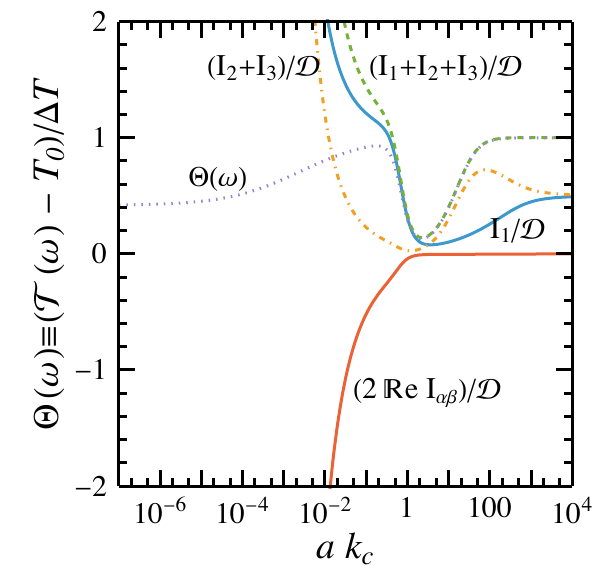}
\caption{Decomposition of $\Theta(\omega)$ for a $1\,\mu{\rm m}$
  particle in water. The low-frequency plateau results from a
  cancellation between the positive self-contributions and the cross
  term. The acoustic minimum, however, is mainly caused by the
  decrease of the self-contribution $(I_1+I_2+I_3)/\mathcal D$ when
  longitudinal dissipation becomes important.}
\label{fig:decomp}
\end{figure}

The mechanism is exposed by the decomposition in
Fig.~\ref{fig:decomp}. The cross term is essential at low frequency
because it cancels part of the positive self-contribution and gives
the Stokes value $5/12$. Near the acoustic crossover the cross
contribution is no longer the main source of the decrease. Instead,
the non-cross contribution $(I_1+I_2+I_3)/\mathcal D$ itself falls
sharply. This occurs because the denominator counts all mechanical
dissipation, including longitudinal dissipation far from the particle,
whereas the numerator weights the same dissipation by the temperature
field $T(r)-T_0\propto1/r$. Acoustic energy dissipated far from the
heated surface therefore contributes strongly to $\mathcal D$ but
weakly to $\mathcal N$.

The nonmonotonic frequency dependence is most naturally understood by
following the dynamics from the short-time, high-frequency side. The
oscillatory motion simultaneously excites transverse and longitudinal
velocity fields. At asymptotically high frequencies, the real and
imaginary parts of the longitudinal wavenumber are comparable,
$u\simeq v$, so that the acoustic wavelength and attenuation length
are both short. The compressional disturbance is then strongly
localized near the heated surface, and the dissipation-weighted
temperature remains relatively high. Upon decreasing the frequency,
the longitudinal attenuation weakens and the acoustic field extends
progressively farther into the colder solvent. The associated shift
of dissipation away from the particle lowers the effective
temperature. The largest relative weighting of the colder fluid is
reached when the acoustic wavelength becomes comparable to the
particle size, $y=ak_c=O(1)$, producing the minimum. At still lower
frequencies, the sound-crossing time becomes short compared with the
oscillation period and pressure equilibrates across the particle
during each cycle. The longitudinal field then ceases to act as an
independent propagating sound mode, and the response crosses over to
the effectively incompressible viscous flow. Dissipation is thereby
weighted more strongly toward the hotter near-surface region, causing
the effective temperature to rise toward its low-frequency
hot-Brownian value.

Two distinct regimes occur on the high-frequency side of the dip. In
the ultimate high-frequency limit, $\ell y\gg 1$, the longitudinal
wave number satisfies $ u\simeq v\simeq\sqrt{y/2\ell}, $ and
both the longitudinal and transverse dissipation are confined close to
the particle surface, giving $1-\Theta\propto x^{-1}$. Upon
decreasing the frequency, the system enters the weakly attenuated
acoustic regime, $\ell y\ll 1$, in which
$ u\simeq y$ and $v\simeq \ell y^2/2$. In the intermediate
interval $ \ell y\ll1, \qquad \ell y^2\gg1, $ or equivalently
$ \ell^{-1/2}\ll y\ll\ell^{-1}, $ the wave is weakly attenuated per
wavelength, $v/u\simeq\ell y/2\ll 1$, while its dissipation remains
localized on a scale smaller than the particle radius, $v\gg 1$. The
corresponding temperature deficit is
$ 1-\Theta\simeq 1/2v \simeq 1 /\ell y^2$. Near the
minimum, $y=\mathcal{O}(1)$ and $v\ll 1$, so that the longitudinal
field extends beyond the particle while its wavelength becomes
comparable to the particle size.

This physical picture separates the mechanisms controlling the
position and the shape of the dip. The acoustic condition $y=O(1)$
sets its limiting position, while finite transverse viscous diffusion
controls the displacement of the minimum from this limit. In
contrast, the width and depth are governed primarily by longitudinal
attenuation, which determines how far the compressional dissipation
extends into the colder fluid before the high-frequency recovery.

We now characterise the dip using the position $y_d$, its width $W$ and
its depth $\Delta_\Theta$, depicted in \cref{fig:theta} in terms of
the material parameters of the colloid and the solvent. We work under the
condition that both $\chi$ and $\ell$ are small that corresponds to
size of the colloid of $1-10 \mu{\rm m}$.  The position of the dip
$y_d$ is set first by the acoustic condition $y=ak_c=\order(1)$, for
which the wavelength of longitudinal sound mode becomes comparable to
the particle size. In the large-particle limit,
$\chi,\ell \rightarrow 0$ while their ratio
$\mathcal{R}_\eta=\chi/\ell$ remains finite. In this limit the
effective
temperature approaches the universal acoustic form
\begin{equation}
  \label{eq:limiting_acoustic form}
\Theta(y)\simeq
\sqrt{2 \chi} (y^4+3y^2+9)/y^{7/2}  
\end{equation}
whose minimum occurs at $y_\infty=((9+\sqrt{333})/2)^{1/2}\approx
3.69$.
The shift of this minimum due to the finite-size of the particle is
controlled by the corrections to this limiting acoustic form. At
$y=\order(1)$, the thickness of the transverse viscous boundary layer
relative to the particle radius is
$\delta_{\nu}/a=x^{-1}=\sqrt{2\chi/y}, $ where
$ \delta_{\nu} = \sqrt{2\nu/\omega}$.  At $y\sim \order(1)$,
$\delta_\nu \sim \sqrt{\chi}$. On the other hand the attenuation of
the longitudinal mode across the particle is measured by
$ v\simeq\ell y^2/2$.

In the asymptotic large-particle regime, the approach to the limiting
position is expected to have the form (at leading logarithmic order)
\begin{equation}
    y_{\infty}-y_d
    \simeq
    A
    \sqrt{\frac{\ell}{\mathcal{R}_{\eta}}}
    \ln\!\left(\frac{1}{\ell}\right)
    =
    \frac{A}{X_y}
    \ln\!\left(\frac{1}{\ell}\right),
\end{equation}
where $A$ is independent of the particle size. Thus, the occurrence of
the dip is determined by the acoustic condition
$y=\order(1)$, whereas its displacement due to finite particle size is controlled
by the combined effects of transverse viscous diffusion and
longitudinal acoustic attenuation.

For $\chi,\ell \ll 1$, the limiting acoustic form follows \cref{eq:limiting_acoustic form}
The width of the dip is primalrily controlled by the attenuation of
the longitudinal mode.  Let $y_{-}$ denote the local maximum preceding
the dip, and $y_+>y_d$ satisfy $\Theta(y_+)=\Theta(y_{-})$.
Near the left shoulder with $y<y_d$, we get $1-\Theta \sim A/x+B
y^{7/2}/\sqrt{\ell}$. Since $x \sim \sqrt{\ell/y}$ we have $1-\Theta
\sim A \sqrt{\ell} y^{-1/2}+B \ell^{-1/2}y^{7/2}$. This gives $y_{-}
\ell^{1/4}$ and $1-\Theta \sim \ell^{3/8}$. On the high frequency side
when $v \gg 1$ but $\ell y \ll 1$ dissipation is localised within a
longitudinal attenuation layer and $1-\Theta \sim 1/v \sim 1/\ell
y^2$. Since we have $\Theta(y_+)=\Theta(y_{-})$, we get
\begin{equation}
  \label{eq:1}
  \frac{1}{\ell y_+^2} \sim \chi^{3/8} \sim \mathcal{R_\eta}^{3/4}\ell^{3/8}
\end{equation}
so that we have $y_+ \sim \mathcal{R_\eta}^{-3/8} \ell^{-11/16}$

\begin{figure}[!ht]
  \centering
  \includegraphics[width=\linewidth]{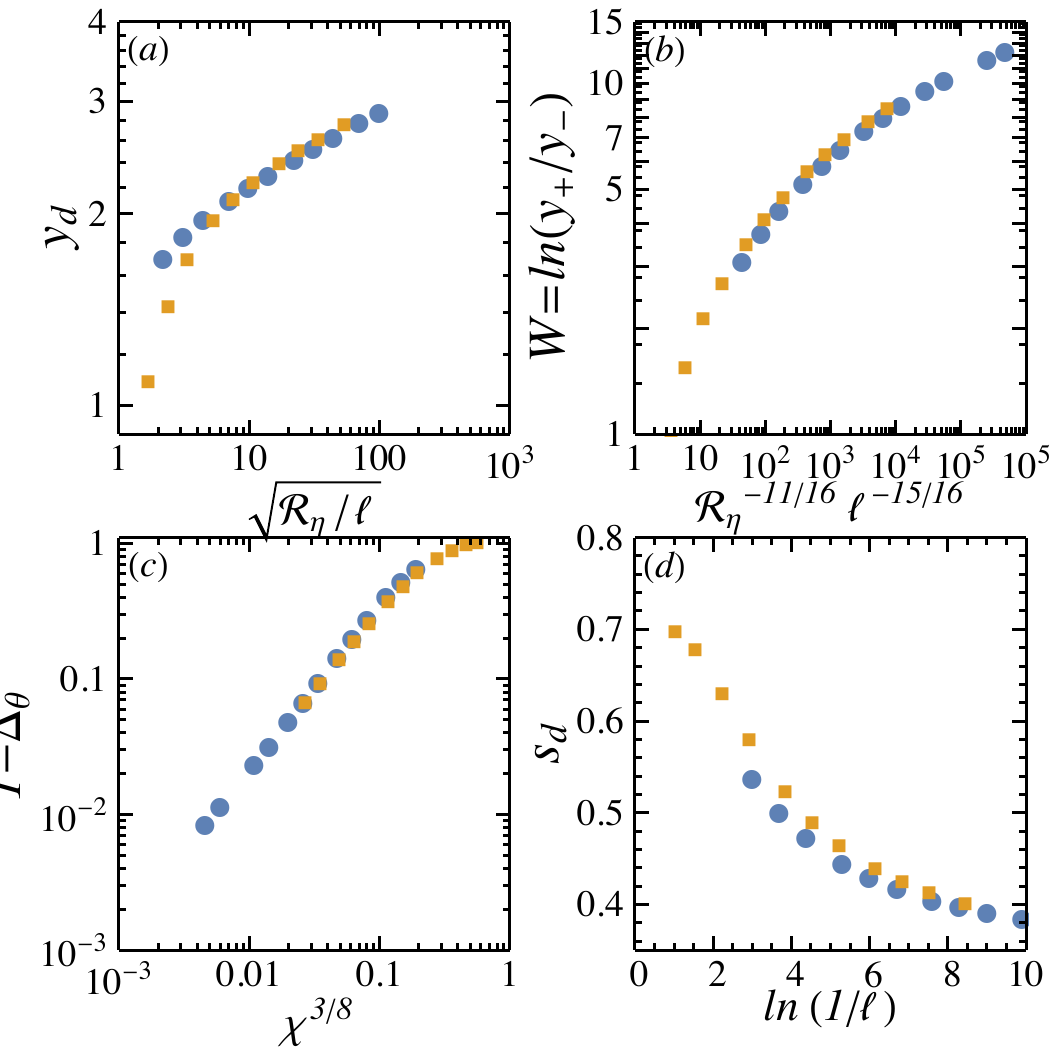}
 \caption{Scaling of the acoustic-dip geometry. The panels show
    (a) the minimum position $y_d$ versus $X_\chi=\sqrt{\mathcal{R}_\eta}{\ell}$,
    (b) the logarithmic width $W=\ln(y_+/y_-)$ versus
    $\ln X_\ell$ with $X_\ell=\mathcal{R}_\eta^{-11/16}\ell^{-15/16}$,
    (c) the depth $1-\Delta_\Theta$ versus the variable
    $\ln X_\Delta=\chi^{3/8}$, and (d) the asymmetry $s_d$ versus
    $\ln X_\ell$ with $X_\ell=\ln (1/\ell)$. Circles denote the hot Brownian motion in a
    Lennard--Jones solvent and the squares denote hot Brownian motion
    in water.}
  \label{fig:min_width_depth}
\end{figure}

Defining the logarithmic
width by $W=\ln\left(\frac{y_+}{y_{-}}\right)$, we expect its leading
dependence is therefore set by the attenuation scale $X_\ell=\ell^{-1/2}$,
with the variation of the low-frequency shoulder providing subleading
corrections. Indeed, a data collpase for $W$ is obtained when plotted
against $\ln X_\ell$ and shows an approximate linear trend with $\ln
X_\ell$ shown in \cref{fig:min_width_depth}~(b).

 The depth is defined by
\begin{equation}
\Delta_\Theta=
\widetilde\Theta-\Theta_{\rm min}=\Theta(y_{-})-\Theta(y_d)
\end{equation}

At $y=y_d$, the value of the effective temperature scales as
$\Theta(y_d) \sim \sqrt{\chi}$ whereas at $y=y_{-}$, the effective
temperature is close to unity. Consequently, we expect that
$\Delta_\Theta$ would scale as $\sqrt{\chi}$. Such a dependence is
shown in \cref{fig:min_width_depth}~(c). Any correction to this
is brought about by $\ell$ through the corrections in $y_{-}$.

The asymmetry of the dip is characterized by the scaled position of
the minimum within its logarithmic width,
$s_d=\ln(y_d/y_{-})/\ln(y_+/y_{-})=W_L/(W_L+W_R)$, where
$W_L=\ln(y_d/y_{-})$ and $W_R=\ln(y_+/y_d)$. Unlike
$y_d$, which is a local property of the acoustic minimum, $s_d$
depends on both edges of the dip. Its variation is
therefore governed predominantly by the attenuation-controlled
expansion of the interval $[y_{-},y_+]$. In particular, the
high-frequency boundary scales as $y_+\sim\ell^{-11/16}$ and the
low-frequency boundary scales as $y_{-} \sim \ell^{1/4}$, while the
minimum remains at $y_d=O(1)$ and undergoes only a weaker
finite-$\chi$ displacement.  The numerical data consequently collapse
approximately when $s_d$ is plotted against $X_\ell=\ln 1/\ell$ as
depicted in \cref{fig:min_width_depth}~(d). This
collapse shows that $s_d$ measures the attenuation-controlled
asymmetry of the dip, rather than the absolute position of
its minimum.

In conclusion, compressibility changes the kinetic effective
temperature of a hot Brownian particle in a way that cannot be inferred
from incompressible hydrodynamics. Sound modes introduce an acoustic
window near $a\omega/c=\order(1)$ in which longitudinal disturbances
carry dissipation into colder regions of the solvent and thereby lower
the effective noise temperature. The limiting position of the minimum
is set by the acoustic time $a/c$, while its finite-size displacement is
controlled mainly by $\chi=\nu/(ac)$. Its width is governed primarily by
the attenuation parameter $\ell=a_c/a$, whereas its depth also retains a
dependence on the viscosity ratio
$\mathcal R_\eta=\eta/(\mu+4\eta/3)$. These conclusions apply within the
continuum, local-equilibrium description with frequency-independent
transport coefficients; at sufficiently high frequencies, molecular
relaxation and viscoelastic dispersion of the longitudinal viscosity
will have to be included.

The strong spectral variation of $\mathcal T(\omega)$ also suggests a
possible route to frequency-selective thermal control. A narrow-band
mechanical mode would predominantly sample the acoustic minimum only if
its response were concentrated near
$\omega_0\simeq(c/a)y_d$. For micron-sized particles in ordinary
liquids, this lies well above conventional optical-trap frequencies and
would instead require coupling to a high-frequency acoustic or
nanomechanical resonator. Two modes centered in different spectral
regions could then experience different noise temperatures while being
coupled to the same nonequilibrium solvent. Such a setup would provide
the ingredients for exploring a frequency-selective Brownian heat
machine, although establishing an operational heat engine would require
an explicit cycle, mode coupling, and work-extraction analysis beyond
the present calculation.

\bibliographystyle{apsrev4-2}
\bibliography{kinetic_temperature}

\cleardoublepage


\section*{Supplementary material}

# Supplementary Materials: Acoustic dip in the effective temperature of hot Brownian motion

Mayank Srivastava[1] and Dipanjan Chakraborty[1, a)]

*Indian Institute of Science Education and Research Mohali, Knowledge City, Sector 81, S. A. S. Nagar, Manauli-140306, India*

Here we present the detailed derivations of the effective temperature $\Theta(\omega)$, the main results of which are presented in the main text.

---

### Appendix A: Dissipation Fucntion

The expression of the dissipation function involves systematic part of the velocity field, and therefore, we require the solution of the Stokes equation with the given boundary condition in **??**. Using the symmetry of the problem, components of the velocity field in spherical polar coordinates are:[1]

$$
\begin{aligned}
v_r = \Delta^{-1}\Bigg[ & 2a^3 P(\omega)\left(\frac{-1}{r^3} + \frac{\mathbb{i}\alpha}{r^2}\right) e^{\mathbb{i}\alpha(r-a)} \\
& + a^3 Q(\omega)\left(\frac{-2}{r^3} + \frac{2\mathbb{i}\beta}{r^2} + \frac{\beta^2}{r}\right) e^{\mathbb{i}\beta(r-a)}\Bigg] V\cos(\theta)
\end{aligned} \tag{A1}
$$

$$
\begin{aligned}
v_\theta = \Delta^{-1}\Bigg[ & a^3 P(\omega)\left(\frac{-1}{r^3} + \frac{\mathbb{i}\alpha}{r^2} + \frac{\alpha^2}{r}\right) e^{\mathbb{i}\alpha(r-a)} \\
& + a^3 Q(\omega)\left(\frac{-1}{r^3} + \frac{\iota\beta}{r^2}\right) e^{\mathbb{i}\beta(r-a)}\Bigg] V\sin(\theta)
\end{aligned} \tag{A2}
$$

The dissipation function in spherical polar coordinates has form

$$
\begin{aligned}
|\Delta|^2\phi(r,\omega) = 2\eta\Bigg[ & (\partial_r v_r)(\partial_r v_r^*) + \left(\frac{\partial_\theta v_\theta}{r} + \frac{v_r}{r}\right)\left(\frac{\partial_\theta v_\theta^*}{r} + \frac{v_r^*}{r}\right) + \left(\frac{v_\theta\cot\theta}{r} + \frac{v_r}{r}\right)\left(\frac{v_\theta^*\cot\theta}{r} + \frac{v_r^*}{r}\right)\Bigg] \\
& + \eta\left[\left(r\partial_r\left(\frac{v_\theta}{r}\right) + \frac{\partial_\theta v_r}{r}\right)\left(r\partial_r\left(\frac{v_\theta^*}{r}\right) + \frac{\partial_\theta v_r^*}{r}\right) - \frac{2}{3}\nabla\cdot\mathbf{v}\nabla\cdot\mathbf{v}^*\right] + \mu\nabla\cdot\mathbf{v}\nabla\cdot\mathbf{v}^*
\end{aligned} \tag{A3}
$$

Using the form of the velocity field in Eqs. (A1) and (A2), the dissipation function in spherical polar coordinates takes the form

$$
\begin{aligned}
|\Delta|^2\phi(r,\omega) = 2\eta\Bigg[ & 6PP^*a^6\frac{e^{\mathbb{i}(\alpha-\alpha^*)(r-a)}}{r^8}g_\alpha g_\alpha^* + QQ^*a^6\frac{e^{\mathbb{i}(\beta-\beta^*)(r-a)}}{r^8}\left(h_\beta h_\beta^* + 2g_\beta g_\beta^*\right) + 2PQ^*a^6\frac{e^{\iota(\alpha-\beta^*)(r-a)}}{r^8}\left(g_\alpha h_\beta^* + g_\alpha g_\beta^*\right) \\
& + 2P^*Qa^6\frac{e^{-\iota(\alpha^*-\beta)(r-a)}}{r^8}\left(g_\alpha^* h_\beta + g_\alpha^* g_\beta\right) - \frac{1}{3}QQ^*a^6\frac{e^{\iota(\beta-\beta^*)(r-a)}}{r^8}\left(|\beta^2|^2r^4 + \mathbb{i}|\beta^2|^2(\beta^*-\beta)r^5 + |\beta^3|^2r^6\right)\Bigg]V^2\cos^2\theta \\
& + \eta\left[PP^*a^6\frac{e^{\iota(\alpha-\alpha^*)(r-a)}}{r^8}h_\alpha h_\alpha^* + 4QQ^*a^6\frac{e^{\iota(\beta-\beta^*)(r-a)}}{r^8}g_\beta g_\beta^* + 2PQ^*a^6\frac{e^{\iota(\alpha-\beta^*)(r-a)}}{r^8}h_\alpha g_\beta^* + 2P^*Qa^6\frac{e^{-\iota(\alpha^*-\beta)(r-a)}}{r^8}h_\alpha^* g_\beta\right]V^2\sin^2\theta \\
& + \mu\left[QQ^*a^6\frac{e^{\mathbb{i}(\beta-\beta^*)(r-a)}}{r^8}\left(|\beta^2|^2r^4 + \mathbb{i}|\beta^2|^2(\beta^*-\beta)r^5 + |\beta^3|^2r^6\right)\right]V^2\cos^2\theta
\end{aligned} \tag{A4}
$$

The most trivial part of this integral is the integration of the solid angle. Once the integration of the solid angle is done, the radial

---

a)Electronic mail: chakraborty@iisermohali.ac.in

integration reads:

$$|\Delta|^2 \int d\mathbf{r}\phi(r,\omega) = \left(\frac{8\pi}{3}\right)\eta V^2 \int_a^\infty r^2 dr\Big[PP^* a^6 \frac{e^{\mathrm{i}(\alpha-\alpha^*)(r-a)}}{r^8}\left(6g_\alpha g_\alpha^* + h_\alpha h_\alpha^*\right) + QQ^* a^6 \frac{e^{\mathrm{i}(\beta-\beta^*)(r-a)}}{r^8}\left(h_\beta h_\beta^* + 6g_\beta g_\beta^*\right)$$
$$+\left(\frac{\mu}{2\eta} - \frac{1}{3}\right)QQ^* a^6 \frac{e^{\mathrm{i}(\beta-\beta^*)(r-a)}}{r^8}(|\beta^2|^2 r^4 + \mathrm{i}|\beta^2|^2(\beta^* - \beta)r^5 + |\beta^3|^2 r^6) + 2PQ^* a^6 \frac{e^{\mathrm{i}(\alpha-\beta^*)(r-a)}}{r^8}(g_\alpha h_\beta^* + g_\alpha g_\beta^* + h_\alpha g_\beta^*)$$
$$+ 2P^* Q a^6 \frac{e^{-\mathrm{i}(\alpha^*-\beta)(r-a)}}{r^8}(g_\alpha^* h_\beta + g_\alpha^* g_\beta + h_\alpha^* g_\beta)\Big] \tag{A5}$$

We now introduce the dimensionless radial coordinate and dimensionless wave numbers $R = \frac{r}{a}, x = ak, u = a\beta_1, v = a\beta_2$ where $\beta = \beta_1 + \mathrm{i}\beta_2$. The radial measure transforms as: $r^2 \mathrm{d}r\, a^6 \frac{1}{r^8} = a^6 \frac{\mathrm{d}r}{r^6} = a\, \frac{\mathrm{d}R}{R^6}$. Therefore the dissipation can be written as

$$|\Delta|^2 \int \phi(\mathbf{r},\omega)\, \mathrm{d}^3 r = \frac{8\pi}{3} a\eta V^2\, \mathcal{D}(\omega), \tag{A6}$$

where $\mathcal{D}(\omega)$ is dimensionless. The coefficients $P$ and $Q$ are $P = 3F_\beta, \quad Q = -3F_\alpha$, with $F_\alpha = 3 - 3\mathrm{i}a\alpha - a^2\alpha^2$, and $F_\beta = 3 - 3\mathrm{i}a\beta - a^2\beta^2$ . Thus $|P|^2 = 9|F_\beta|^2$, $|Q|^2 = 9|F_\alpha|^2$,$PQ^* = -9F_\beta F_\alpha^*$, and $P^* Q = -9F_\beta^* F_\alpha$. The squared amplitudes are $|F_\alpha|^2 = 4x^4 + 12x^3 + 18x^2 + 18x + 9$, and $|F_\beta|^2 = u^4 + 2u^2v^2 + v^4 + 6u^2v + 6v^3 + 3u^2 + 15v^2 + 18v + 9$.

We also define

$$J_n(q) = \int_1^\infty R^{-n} e^{-q(R-1)}\, \mathrm{d}R, \tag{A7}$$

which satisfies

$$J_{n+1}(q) = \frac{1 - qJ_n(q)}{n}. \tag{A8}$$

The transverse self contribution is

$$\mathcal{D}_\alpha = |P|^2 \int_1^\infty \mathrm{d}R\, \frac{e^{-2x(R-1)}}{R^6}\, (6g_\alpha g_\alpha^* + h_\alpha h_\alpha^*)\,. \tag{A9}$$

Using $6g_\alpha g_\alpha^* + h_\alpha h_\alpha^* = 90 + 180xR + 180x^2R^2 + 120x^3R^3 + 60x^4R^4 + 24x^5R^5 + 8x^6R^6$, we obtain

$$\mathcal{D}_\alpha = 9|F_\beta|^2\Big[90J_6(2x) + 180xJ_5(2x) + 180x^2J_4(2x) + 120x^3J_3(2x) + 60x^4J_2(2x) + 24x^5J_1(2x) + 8x^6J_0(2x)\Big]. \tag{A10}$$

Using the recurrence relation for $J_n$, this reduces to

$$\boxed{\mathcal{D}_\alpha = |F_\beta|^2\left[36x^4(1+x) + 18|F_\alpha|^2\right].} \tag{A11}$$

The longitudinal self term is

$$\mathcal{D}_{\beta,1} = |Q|^2 \int_1^\infty \mathrm{d}R\, \frac{e^{-2v(R-1)}}{R^6}\left(h_\beta h_\beta^* + 6g_\beta g_\beta^*\right). \tag{A12}$$

Define $P_2(R;u,v) = h_\beta(R)h_\beta^*(R) + 6g_\beta(R)g_\beta^*(R)$ where $g_\beta(R) = -3 + 3\mathrm{i}(u+\mathrm{i}v)R + (u+\mathrm{i}v)^2R^2$ and $h_\beta(R) = -6 + 6\mathrm{i}(u+\mathrm{i}v)R + 3(u+\mathrm{i}v)^2R^2 - \mathrm{i}(u+\mathrm{i}v)^3R^3$. Expanding the product gives $P_2(R;u,v) = 90 + 180vR + \left(18u^2 + 162v^2\right)R^2 + \left(36u^2v + 84v^3\right)R^3 + \left(3u^4 + 30u^2v^2 + 27v^4\right)R^4 + \left(6u^4v + 12u^2v^3 + 6v^5\right)R^5 + \left(u^2+v^2\right)^3 R^6$.

The divergence correction is

$$\mathcal{D}_{\beta,2} = \left(\frac{\mu}{2\eta} - \frac{1}{3}\right)|Q|^2 \int_1^\infty \mathrm{d}R\, \frac{e^{-2v(R-1)}}{R^6} P_3(R;u,v), \tag{A13}$$

where $P_3(R;u,v) = (u^2+v^2)^2R^4 + 2v(u^2+v^2)^2R^5 + (u^2+v^2)^3R^6$. Combining these two terms gives

$$\mathcal{D}_\beta = \mathcal{D}_{\beta,1} + \mathcal{D}_{\beta,2}. \tag{A14}$$

Using $|Q|^2 = 9|F_\alpha|^2$, the longitudinal contribution can be written as

$$\mathcal{D}_\beta = 9|F_\alpha|^2 \int_1^\infty \mathrm{d}R\, R^{-6} e^{-2v(R-1)} \left[ P_2(R; u, v) + \left( \frac{\mu}{2\eta} - \frac{1}{3} \right) P_3(R; u, v) \right]. \tag{A15}$$

Defining $\widetilde{P}_\beta(R; u, v) = P_2(R; u, v) + \left(\frac{\mu}{2\eta} - \frac{1}{3}\right) P_3(R; u, v)$, we have $\widetilde{P}_\beta(R; u, v) = 90 + 180vR + \left(18u^2 + 162v^2\right) R^2 + \left(36u^2 v + 84v^3\right) R^3 + \left((3 + \kappa)u^4 + (30 + 2\kappa)u^2 v^2 + (27 + \kappa)v^4\right) R^4 + (6 + 2\kappa)v \left(u^2 + v^2\right)^2 R^5 + (1 + \kappa) \left(u^2 + v^2\right)^3 R^6$, where $\kappa = (\mu/2\eta - 1/3)$. Now define $J_n(2v) = \int_1^\infty \mathrm{d}R\, R^{-n} e^{-2v(R-1)}$, and since the radial weight is $R^{-6}$, the term proportional to $R^m$ gives $J_{6-m}$. Hence we get

$$\frac{\mathcal{D}_\beta}{9|F_\alpha|^2} = 90J_6 + 180vJ_5 + \left(18u^2 + 162v^2\right) J_4 + \left(36u^2 v + 84v^3\right) J_3 + \left((3 + \kappa)u^4 + (30 + 2\kappa)u^2 v^2 + (27 + \kappa)v^4\right) J_2 \\ + (6 + 2\kappa)v \left(u^2 + v^2\right)^2 J_1 + (1 + \kappa) \left(u^2 + v^2\right)^3 J_0. \tag{A16}$$

All the $J_n$'s in this expression have argument $2v$. The recurrence relation is $J_{n+1} = \frac{1-2vJ_n}{n}$. The $J_0$ integral is $J_0 = \int_1^\infty e^{-2v(R-1)}\, \mathrm{d}R = \frac{1}{2v}$. The required reductions are $J_2 = 1-2vJ_1$, $J_3 = \frac{1}{2}-v+2v^2 J_1$, $J_4 = \frac{1}{3}-\frac{v}{3}+\frac{2v^2}{3}-\frac{4v^3}{3}J_1$, $J_5 = \frac{1}{4}-\frac{v}{6}+\frac{v^2}{6}-\frac{v^3}{3}+\frac{2v^4}{3}J_1$, and $J_6 = \frac{1}{5} - \frac{v}{10} + \frac{v^2}{15} - \frac{v^3}{15} + \frac{2v^4}{15} - \frac{4v^5}{15} J_1$

This yields

$$\begin{aligned} \frac{\mathcal{D}_\beta}{9|F_\alpha|^2} &= 2|F_\beta|^2 + (1 + \kappa)(u^2 + v^2)^2 \left(1 + \frac{u^2 + v^2}{2v}\right) \\ &= 2|F_\beta|^2 + \left(\frac{1}{3} + \frac{\mu}{4\eta}\right)(u^2 + v^2)^2 \left(1 + \frac{u^2 + v^2}{2v}\right) \\ &= 2|F_\beta|^2 + \frac{1}{4\mathcal{R}_\eta}(u^2 + v^2)^2 \left(\frac{u^2 + v^2 + 2v}{v}\right) \end{aligned} \tag{A17}$$

After carrying out the radial integrals, one obtains

$$\boxed{\mathcal{D}_\beta = 9|F_\alpha|^2 \left[ \frac{1}{4\mathcal{R}_\eta} \frac{(u^2 + v^2)^2 (u^2 + v^2 + 2v)}{v} + 2|F_\beta|^2 \right].} \tag{A18}$$

The cross contribution is

$$\mathcal{D}_{\alpha\beta} = 2PQ^* \int_1^\infty \mathrm{d}R\, R^{-6} e^{-s_{\alpha\beta}(R-1)} \left(g_\alpha h_\beta^* + g_\alpha g_\beta^* + h_\alpha g_\beta^*\right) + 2P^* Q \int_1^\infty \mathrm{d}R\, R^{-6} e^{-s_{\beta\alpha}(R-1)} \left(g_\alpha^* h_\beta + g_\alpha^* g_\beta + h_\alpha^* g_\beta\right), \tag{A19}$$

where $s_{\alpha\beta} = -\mathbb{i}(a\alpha - a\beta^*)$, and $s_{\beta\alpha} = -\mathbb{i}(a\beta - a\alpha^*)$ . Consider the first integral,

$$\mathcal{K}_{\alpha\beta} = \int_1^\infty \mathrm{d}R\, R^{-6} e^{-s_{\alpha\beta}(R-1)} \left(g_\alpha h_\beta^* + g_\alpha g_\beta^* + h_\alpha g_\beta^*\right), \tag{A20}$$

Since $g_\alpha(R) = -3 + 3\mathbb{i}\hat{\alpha}R + \hat{\alpha}^2 R^2$, $h_\alpha(R) = -6 + 6\mathbb{i}\hat{\alpha}R + 3\hat{\alpha}^2 R^2 - \mathbb{i}\hat{\alpha}^3 R^3$, $g_\beta^*(R) = -3 - 3\mathbb{i}\hat{\beta}^* R + (\hat{\beta}^*)^2 R^2$, $h_\beta^*(R) = -6 - 6\mathbb{i}\hat{\beta}^* R + 3(\hat{\beta}^2)^* R^2 + \mathbb{i}(\hat{\beta}^3)^* R^3$ , the polynomial in the $\mathcal{K}_{\alpha\beta}$ takes the form $g_\alpha h_\beta^* + g_\alpha g_\beta^* + h_\alpha g_\beta^* = \sum_{m=0}^5 C_m^{\alpha\beta} R^m$. Expanding the polynomial explicitly we get

$$\sum_{m=0}^{5} C_m^{\alpha\beta} R^m = 45 + 45\mathbb{i}(\hat{\beta}^* - \hat{\alpha})R + \left(-18\hat{\alpha}^2 + 45\hat{\alpha}\hat{\beta}^* - 18(\hat{\beta}^*)^2\right) R^2 + 3\mathbb{i} \left(\hat{\alpha}^3 - 6\hat{\alpha}^2\hat{\beta}^* + 6\hat{\alpha}(\hat{\beta}^*)^2 - (\hat{\beta}^*)^3\right) R^3 \\ + \left(-3\hat{\alpha}^3\hat{\beta}^* + 7\hat{\alpha}^2(\hat{\beta}^*)^2 - 3\hat{\alpha}(\hat{\beta}^*)^3\right) R^4 + \mathbb{i} \left(-\hat{\alpha}^3(\hat{\beta}^*)^2 + \hat{\alpha}^2(\hat{\beta}^*)^3\right) R^5. \tag{A21}$$

It follows that $\mathcal{K}_{\alpha\beta} = \sum_{m=0}^5 C_m^{\alpha\beta} J_{6-m}(s_{\alpha\beta})$. Using the recurrence relation for $J_n(s)$ and $s = s_{\alpha\beta} = -\mathbb{i}(\hat{\alpha}-\hat{\beta}^*)$, all terms proportional to the remaining exponential integral $J_1(s_{\alpha\beta})$ cancel. The remaining polynomial reduces to

$$\mathcal{K}_{\alpha\beta} = \left(\hat{\alpha}^2 + 3\mathbb{i}\hat{\alpha} - 3\right)\left((\hat{\beta}^*)^2 - 3\mathbb{i}\hat{\beta}^* - 3\right) = F_\alpha F_\beta^*. \tag{A22}$$

Therefore, the first radial integral evaluates to

$$\int_1^\infty \mathrm{d}R\, R^{-6} e^{-s_{\alpha\beta}(R-1)} \left( g_\alpha h_\beta^* + g_\alpha g_\beta^* + h_\alpha g_\beta^* \right) = F_\alpha F_\beta^*, \tag{A23}$$

and the second is its complex conjugate,

$$\int_1^\infty \mathrm{d}R\, R^{-6} e^{-s_{\beta\alpha}(R-1)} \left( g_\alpha^* h_\beta + g_\alpha^* g_\beta + h_\alpha^* g_\beta \right) = F_\alpha^* F_\beta. \tag{A24}$$

Since$PQ^* = -9F_\beta F_\alpha^*$, and $P^*Q = -9F_\beta^* F_\alpha$, the cross contribution becomes

$$\boxed{\mathcal{D}_{\alpha\beta} = -36|F_\alpha|^2|F_\beta|^2.} \tag{A25}$$

The total dimensionless dissipation is therefore $\mathcal{D}(\omega) = \mathcal{D}_\alpha + \mathcal{D}_\beta + \mathcal{D}_{\alpha\beta}$. Substituting the three contributions gives

$$\mathcal{D}(\omega) = |F_\beta|^2 \left[36x^4(1+x) + 18|F_\alpha|^2\right] + 9|F_\alpha|^2 \left[\frac{1}{4\mathcal{R}_\eta} \frac{(u^2+v^2)^2(u^2+v^2+2v)}{v} + 2|F_\beta|^2\right] - 36|F_\alpha|^2|F_\beta|^2. \tag{A26}$$

The product terms cancel exactly, and we have

$$\boxed{|\Delta|^2 \int \phi(\mathbf{r},\omega)\, \mathrm{d}^3 r = \frac{8\pi}{3} a\eta V^2 \left[9|F_\alpha|^2 \frac{1}{4\mathcal{R}_\eta} \frac{(u^2+v^2)^2(u^2+v^2+2v)}{v} + 36x^4(1+x)|F_\beta|^2\right].} \tag{A27}$$

**Appendix B: The numerator in the $\mathcal{T}(\omega)$**

$$\begin{aligned}|\Delta|^2 \int d\mathbf{r} \frac{a\phi(r,\omega)}{r} = \frac{8\pi a}{3}\eta V^2 \int_a^\infty r^2 dr \Big[ & PP^* a^6 \frac{e^{\mathbb{i}(\alpha-\alpha^*)(r-a)}}{r^9} (6g_\alpha g_\alpha^* + h_\alpha h_\alpha^*) + QQ^* a^6 \frac{e^{\mathbb{i}(\beta-\beta^*)(r-a)}}{r^9} (h_\beta h_\beta^* + 6g_\beta g_\beta^*) \\ & + \left(\frac{\mu}{2\eta} - \frac{1}{3}\right) QQ^* a^6 \frac{e^{\mathbb{i}(\beta-\beta^*)(r-a)}}{r^9} (|\beta^2|^2 r^4 + \mathbb{i}|\beta^2|^2(\beta^* - \beta) r^5 + |\beta^3|^2 r^6) + 2PQ^* a^6 \frac{e^{\mathbb{i}(\alpha-\beta^*)(r-a)}}{r^9} (g_\alpha h_\beta^* + g_\alpha g_\beta^* + h_\alpha g_\beta^*) \\ & + 2P^* Q a^6 \frac{e^{-\mathbb{i}(\alpha^*-\beta)(r-a)}}{r^9} (g_\alpha^* h_\beta + g_\alpha^* g_\beta + h_\alpha^* g_\beta) \Big]\end{aligned} \tag{B1}$$

With this the first term in the integral in Eq. (B1) as

$$I_1(\omega) = \int_a^\infty r^2 dr \left[ PP^* a^6 \frac{e^{\mathbb{i}(\alpha-\alpha^*)(r-a)}}{r^9} \left(6g_\alpha g_\alpha^* + h_\alpha h_\alpha^*\right)\right] = 9|F_\beta|^2 a^6 \int_a^\infty \mathrm{d}r \frac{e^{\mathbb{i}(\alpha-\alpha^*)(r-a)}}{r^7} \left[6g_\alpha(r,\omega) g_\alpha^*(r,\omega) + h_\alpha(r,\omega) h_\alpha^*(r,\omega)\right] \tag{B2}$$

Similary we have

$$I_2(\omega) = \int_a^\infty r^2 \mathrm{d}r \left[ QQ^* a^6 \frac{e^{\mathbb{i}(\beta-\beta^*)(r-a)}}{r^9} \left(6g_\beta g_\beta^* + h_\beta h_\beta^*\right)\right] = 9|F_\alpha|^2 a^6 \int_a^\infty \mathrm{d}r \frac{e^{\mathbb{i}(\beta-\beta^*)(r-a)}}{r^7} \left[6g_\beta(r,\omega) g_\beta^*(r,\omega) + h_\beta(r,\omega) h_\beta^*(r,\omega)\right] \tag{B3}$$

$$\begin{aligned} I_3(\omega) &= \left(\frac{\mu}{2\eta} - \frac{1}{3}\right) QQ^* \int_a^\infty r^2 \mathrm{d}r a^6 \frac{e^{\mathbb{i}(\beta-\beta^*)(r-a)}}{r^9} \left(|\beta^2|^2 r^4 + \mathbb{i}|\beta^2|^2(\beta^* - \beta) r^5 + |\beta^3|^2 r^6\right) \\ &= 9\left(\frac{\mu}{2\eta} - \frac{1}{3}\right) |F_\alpha|^2 a^6 \int_a^\infty \mathrm{d}r \frac{e^{\mathbb{i}(\beta-\beta^*)(r-a)}}{r^7} \left(|\beta^2|^2 r^4 + \mathbb{i}|\beta^2|^2(\beta^* - \beta) r^5 + |\beta^3|^2 r^6\right) \end{aligned} \tag{B4}$$

$$I_4(\omega) = 2\int_a^\infty r^2 \mathrm{d}r PQ^* a^6 \frac{e^{\mathbb{i}(\alpha-\beta^*)(r-a)}}{r^9} \left(g_\alpha h_\beta^* + g_\alpha g_\beta^* + h_\alpha g_\beta^*\right) \tag{B5}$$

and

$$I_5(\omega) = 2\int_a^\infty r^2 \mathrm{d}r P^* Q a^6 \frac{e^{-\mathbb{i}(\alpha^*-\beta)(r-a)}}{r^9}\left(g_\alpha^* h_\beta + g_\alpha^* g_\beta + h_\alpha^* g_\beta\right). \tag{B6}$$

The fourth and the fifth term in Eq. (B5) and Eq. (B6) are complex conjugates of each other: $I_5 = I_4^*$ so that their sum is $I_4 + I_5 = 2\,\mathbb{R}\mathrm{e}\, I_{\alpha\beta}$ with

$$I_{\alpha\beta}(\omega) = -18 F_\beta F_\alpha^* a^6 \int_a^\infty \mathrm{d}r \frac{e^{\mathbb{i}(\alpha-\beta^*)(r-a)}}{r^7}\left[g_\alpha(r) h_\beta^*(r) + g_\alpha g_\beta^* + h_\alpha g_\beta^*\right]. \tag{B7}$$

The last term is the complex conjugate $I_{\alpha\beta}$ so that the numerator is

$$\mathcal{N}_{\text{num}}(\omega) \equiv \mathcal{N}(y,\chi,\ell) = I_1(\omega) + I_2(\omega) + I_3(\omega) + 2\,\mathbb{R}\mathrm{e}\, I_{\alpha\beta}(\omega). \tag{B8}$$

1 We now evaluate the radial integral using dimensionless variable $R = r/a$, for each of the terms $I_1, I_2, I_3$ and $I_{\alpha\beta}$. For $I_1$ we have,

$$\begin{aligned} I_1(\omega) &= 9|F_\beta|^2 a^6 \int_a^\infty \mathrm{d}r \frac{e^{\mathbb{i}(\alpha-\alpha^*)(r-a)}}{r^7}\left[6 g_\alpha(r,\omega) g_\alpha^*(r,\omega) + h_\alpha(r,\omega) h_\alpha^*(r,\omega)\right] \\ &= 9|F_\beta|^2 a^6 \int_a^\infty \mathrm{d}r \frac{e^{\mathbb{i}(\alpha-\alpha^*)(r-a)}}{r^7}\left[90 + 180kr + 180k^2r^2 + 120k^3r^3 + 60k^4r^4 + 24k^5r^5 + 8k^6r^6\right] \\ &= 9|F_\beta|^2 a^6 \int_1^\infty a\mathrm{d}R \frac{e^{a\mathbb{i}(\alpha-\alpha^*)(R-1)}}{a^7R^7}\left[90 + 180kaR + 180k^2a^2R^2 + 120k^3a^3R^3 + 60k^4a^4R^4 + 24k^5a^5R^5 + 8k^6a^6R^6\right] \\ &= 9A(x)|F_\beta|^2 \end{aligned} \tag{B9}$$

In arriving at the final form in Eq. (B9), we have used the fact that the integral has terms like $\int_1^\infty \mathrm{d}R R^{-n} e^{-2x(R-1)}$. We define

$$J_n(q) = \int_1^\infty R^{-n} e^{-q(R-1)}\,\mathrm{d}R. \tag{B10}$$

For the transverse contribution $I_1$, we have

$$q = 2x, \tag{B11}$$

and

$$\frac{I_1}{9|F_\beta|^2} = 90J_7 + 180xJ_6 + 180x^2J_5 + 120x^3J_4 + 60x^4J_3 + 24x^5J_2 + 8x^6J_1. \tag{B12}$$

The recurrence relation follows by integration by parts:

$$J_{n+1}(q) = \frac{1 - qJ_n(q)}{n}. \tag{B13}$$

With $q = 2x$, we have $J_{n+1} = \frac{1-2xJ_n}{n}$. Thus, $J_2 = 1-2xJ_1$,$J_3 = \frac{1}{2}-x+2x^2J_1$, $J_4 = \frac{1}{3}-\frac{x}{3}+\frac{2x^2}{3}-\frac{4x^3}{3}J_1$, $J_5 = \frac{1}{4}-\frac{x}{6}+\frac{x^2}{6}-\frac{x^3}{3}+\frac{2x^4}{3}J_1$,$J_6 = \frac{1}{5} - \frac{x}{10} + \frac{x^2}{15} - \frac{x^3}{15} + \frac{2x^4}{15} - \frac{4x^5}{15}J_1$ and $J_7 = \frac{1}{6} - \frac{x}{15} + \frac{x^2}{30} - \frac{x^3}{45} + \frac{x^4}{45} - \frac{2x^5}{45} + \frac{4x^6}{45}J_1$.
Substituting these expressions into

$$S_1 = 90J_7 + 180xJ_6 + 180x^2J_5 + 120x^3J_4 + 60x^4J_3 + 24x^5J_2 + 8x^6J_1, \tag{B14}$$

we obtain

$$\begin{aligned} 90J_7 &= 15 - 6x + 3x^2 - 2x^3 + 2x^4 - 4x^5 + 8x^6J_1, \\ 180xJ_6 &= 36x - 18x^2 + 12x^3 - 12x^4 + 24x^5 - 48x^6J_1, \\ 180x^2J_5 &= 45x^2 - 30x^3 + 30x^4 - 60x^5 + 120x^6J_1, \\ 120x^3J_4 &= 40x^3 - 40x^4 + 80x^5 - 160x^6J_1, \\ 60x^4J_3 &= 30x^4 - 60x^5 + 120x^6J_1, \\ 24x^5J_2 &= 24x^5 - 48x^6J_1, \\ 8x^6J_1 &= 8x^6J_1 \end{aligned} \tag{B15}$$

Collecting the polynomial terms, the coefficient of $J_1$ vanishes exactly, and we get

$$\frac{I_1}{9|F_\beta|^2} = 4x^5 + 10x^4 + 20x^3 + 30x^2 + 30x + 15 \tag{B16}$$

Thus, we have the final form

$$I_1 = 9A(x)|F_\beta|^2. \tag{B17}$$

For $I_2(\omega)$ we have:

$$\begin{aligned}
I_2(\omega) &= 9|F_\alpha|^2 a^6 \int_a^\infty \mathrm{d}r \frac{e^{\mathbb{i}(\beta-\beta^*)(r-a)}}{r^7}\left[6g_\beta(r,\omega)g^*_\beta(r,\omega) + h_\beta(r,\omega)h^*_\beta(r,\omega)\right] \\
&= 9|F_\alpha|^2 a^6 \int_a^\infty \mathrm{d}r \frac{e^{\mathbb{i}(\beta-\beta^*)(r-a)}}{r^7}\Big[90 + 180\beta_2 r + \left(18\beta_1^2 + 162\beta_2^2\right) r^2 + \left(36\beta_1^2\beta_2 + 84\beta_2^3\right) r^3 \\
&+ \left(3\beta_1^4 + 30\beta_1^2\beta_2^2 + 27\beta_2^4\right) r^4 + \left(6\beta_1^4\beta_2 + 12\beta_1^2\beta_2^3 + 6\beta_2^5\right) r^5 + \left(\beta_1^2 + \beta_2^2\right)^3 r^6 \\
&= 9|F_\alpha|^2 a^6 \int_1^\infty a\mathrm{d}R \frac{e^{a\mathbb{i}(\beta-\beta^*)(R-1)}}{a^7R^7}\Big[90 + 180a\beta_2 R + \left(18\beta_1^2 + 162\beta_2^2\right) a^2R^2 + \left(36\beta_1^2\beta_2 + 84\beta_2^3\right) a^3R^3 \\
&+ \left(3\beta_1^4 + 30\beta_1^2\beta_2^2 + 27\beta_2^4\right) a^4R^4 + \left(6\beta_1^4\beta_2 + 12\beta_1^2\beta_2^3 + 6\beta_2^5\right) a^5R^5 + \left(\beta_1^2 + \beta_2^2\right)^3 a^6R^6 \\
&= 9|F_\alpha|^2 \int_1^\infty \mathrm{d}R \frac{e^{a\mathbb{i}(\beta-\beta^*)(R-1)}}{R^7}\Big[90 + 180vR + \left(18u^2 + 162v^2\right) R^2 + \left(36u^2v + 84v^3\right) R^3 \\
&+ \left(3u^4 + 30u^2v^2 + 27v^4\right) R^4 + \left(6u^4v + 12u^2v^3 + 6v^5\right) R^5 + \left(u^2 + v^2\right)^3 R^6 \\
&= \frac{9}{2}|F_\alpha|^2 D(u,v)
\end{aligned} \tag{B18}$$

where $D(u,v) = -2\Lambda u^6 + 6\Lambda u^4v^2 - 6\Lambda u^2v^4 + 2\Lambda v^6 + 6u^4v + 3u^4 + 12u^2v^2 + 18u^2v + 9u^2 + 2v^5 + 5v^4 + 22v^3 + 51v^2 + 60v + 30$.

For $I_3(\omega)$ we have

$$\begin{aligned}
I_3(\omega) &= 9\left(\frac{\mu}{2\eta} - \frac{1}{3}\right)|F_\alpha|^2 a^6 \int_a^\infty \mathrm{d}r \frac{e^{\mathbb{i}(\beta-\beta^*)(r-a)}}{r^7}\left(|\beta^2|^2 r^4 + \mathbb{i}|\beta^2|^2(\beta^* - \beta) r^5 + |\beta^3|^2 r^6\right) \\
&= 9\left(\frac{\mu}{2\eta} - \frac{1}{3}\right)|F_\alpha|^2 a^6 \int_a^\infty \mathrm{d}R \frac{e^{\mathbb{i}(\beta-\beta^*)(r-a)}}{r^7}\left(\left(\beta_1^2 + \beta_2^2\right)^2 r^4 + 2\beta_2\left(\beta_1^2 + \beta_2^2\right)^2 r^5 + \left(\beta_1^2 + \beta_2^2\right)^3 r^6\right) \\
&= 9\left(\frac{\mu}{2\eta} - \frac{1}{3}\right)|F_\alpha|^2 a^6 \int_a^\infty \mathrm{d}r\, e^{-2\beta_2(r-a)}\left[\frac{\beta_1^4 + 2\beta_1^2\beta_2^2 + \beta_2^4}{r^3} + \frac{2\beta_1^4\beta_2 + 4\beta_1^2\beta_2^3 + 2\beta_2^5}{r^2} + \frac{\beta_1^6 + 3\beta_1^4\beta_2^2 + 3\beta_1^2\beta_2^4 + \beta_2^6}{r}\right] \\
&= 9\left(\frac{\mu}{2\eta} - \frac{1}{3}\right)|F_\alpha|^2 \int_1^\infty \mathrm{d}R\, R^{-7}\, e^{-2v(R-1)}\left(u^4 + 2u^2v^2 + v^4\right) R^4 + \left(2u^4v + 4u^2v^3 + 2v^5\right) R^5 + \left(u^6 + 3u^4v^2 + 3u^2v^4 + v^6\right) R^6 \\
&= \left(\frac{9\mu}{4\eta} - \frac{3}{2}\right)|F_\alpha|^2\left(u^2 + v^2\right)^2\left[2\Lambda(v^2 - u^2) + 2v + 1\right] = \frac{9}{2}\kappa|F_\alpha|^2\left(u^2 + v^2\right)^2\left[2\Lambda(v^2 - u^2) + 2v + 1\right]
\end{aligned} \tag{B19}$$

It now remains to evaluate the integral in $I_{\alpha\beta}(\omega)$. For this define the dimensionless variable $\hat{\alpha}(y,\chi) = a\alpha(y,\chi)$, $\hat{\beta}(y,\ell) = a\beta(y,\ell)$ and $s_{\alpha\beta}(y,\chi,\ell) = -\mathbb{i}(\hat{\alpha} - \hat{\beta}^*)$. In the dimensionless coordinate $R = r/a$, we have $e^{\mathbb{i}(\alpha-\beta^*)(r-a)} = e^{-s_{\alpha\beta}(R-1)}$. The polynomial in the square bracket in Eq. (B7) can be expanded as

$$g_\alpha h^*_\beta + g_\alpha g^*_\beta + h_\alpha g^*_\beta = \sum_{m=0}^{5} C_m^{\alpha\beta}(y,\chi,\ell)R^m, \tag{B20}$$

where $C_m^{\alpha\beta}$ are functions of $\omega$.

$$\begin{aligned}
C_0^{\alpha\beta} &= 45 \\
C_1^{\alpha\beta} &= -45\mathbb{i}(\hat{\alpha} - \hat{\beta}^*) \\
C_2^{\alpha\beta} &= -9(\hat{\alpha} - 2\hat{\beta}^*)(2\hat{\alpha} - \hat{\beta}^*) \\
C_3^{\alpha\beta} &= 3\mathbb{i}(\hat{\alpha} - \hat{\beta}^*)(\hat{\alpha}^2 - 5\hat{\alpha}\hat{\beta}^* + \hat{\beta}^{*2}) \\
C_4^{\alpha\beta} &= -\hat{\alpha}\hat{\beta}^*(3\hat{\alpha}^2 - 7\hat{\alpha}\hat{\beta}^* + 3\hat{\beta}^{*2}) \\
C_5^{\alpha\beta} &= -\mathbb{i}\hat{\alpha}^2\hat{\beta}^{*2}(\hat{\alpha} - \hat{\beta}^*)
\end{aligned} \tag{B21}$$

Therefore, we have

$$I_{\alpha\beta}(y, \chi, \ell) = -18F_\beta F_\alpha^* \int_1^\infty \mathrm{d}RR^{-7}e^{-s_{\alpha\beta}(R-1)} \sum_{m=0}^{5} C_m^{\alpha\beta} R^m \tag{B22}$$

which upon completing the radial integral gives

$$I_{\alpha\beta}(y, \chi, \ell) = -18F_\beta F_\alpha^* e^{s_{\alpha\beta}} \sum_{m=0}^{5} C_m^{\alpha\beta} E_{7-m}(s_{\alpha\beta}) \tag{B23}$$

with $E_n(z) = \int_1^\infty \mathrm{d}x x^{-n} e^{-zx}$. Consequently, the sum of the last two terms in the integral Eq. (B1) is given by $2\,\mathbb{R}\mathrm{e}\, I_{\alpha\beta}(\omega) = -36\,\mathbb{R}\mathrm{e}\left[F_\beta F_\alpha^* e^{s_{\alpha\beta}} \sum_{m=0}^5 C_m^{\alpha\beta} E_{7-m}(s_{\alpha\beta})\right]$.

The complete the numerator then reads

$$\mathcal{N}(\omega) = 9A(x)|F_\beta|^2 + \frac{9}{2}|F_\alpha|^2\left[D(u,v) + \kappa(u^2+v^2)^2\left(2\Lambda(v)(v^2-u^2) + 2v + 1\right)\right] + 2\,\mathbb{R}\mathrm{e}\, I_{\alpha\beta} \tag{B24}$$

### Appendix C: Large-particle limit of the acoustic minimum

We now determine the limiting position of the acoustic minimum as the particle radius becomes large. The dimensionless frequency is

$$y = ak_c = \frac{a\omega}{c}. \tag{C1}$$

For a fixed solvent,

$$\chi = \frac{\nu}{ac} \propto a^{-1}, \qquad \ell = \frac{a_c}{a} \propto a^{-1}, \tag{C2}$$

whereas their ratio

$$\mathcal{R}_\eta = \frac{\chi}{\ell} = \frac{\nu}{ca_c} = \frac{\eta}{\eta_L} \tag{C3}$$

is independent of the particle radius. Thus, in the limit $a \to \infty$ at fixed $y = O(1)$,

$$x = ak = \sqrt{\frac{y}{2\chi}} = \sqrt{\frac{y}{2\mathcal{R}_\eta \ell}} \longrightarrow \infty. \tag{C4}$$

At the same time, the longitudinal wavenumber $a\beta = u + iv$ approaches the weakly attenuated acoustic limit,

$$u \simeq y, \qquad v \simeq \frac{\ell y^2}{2} \longrightarrow 0. \tag{C5}$$

The effective temperature can be written as

$$\Theta(y) = \frac{\mathcal{N}(y)}{\mathcal{D}(y)}, \tag{C6}$$

where $\mathcal{N}$ is the temperature-weighted dissipation and $\mathcal{D}$ is the total dissipation. In the large-$x$, small-$v$ limit, the leading contribution to the numerator is

$$\mathcal{N}(y) \simeq 36x^5\left(y^4 + 3y^2 + 9\right), \tag{C7}$$

whereas the dominant contribution to the denominator is the longitudinal term singular in the attenuation parameter,

$$\mathcal{D}(y) \simeq \frac{9}{\mathcal{R}_\eta}\frac{x^4y^6}{v}. \tag{C8}$$

Terms containing lower powers of $x$, finite-$v$ corrections, and the normalized transverse–longitudinal cross contribution are subleading in this limit. Consequently,

$$\begin{aligned}\Theta(y) &\simeq \frac{36x^5\left(y^4+3y^2+9\right)}{9x^4y^6/\mathcal{R}_\eta v}\\ &= 4\mathcal{R}_\eta xv\,\frac{y^4+3y^2+9}{y^6}.\end{aligned} \tag{C9}$$

Using

$$x = \sqrt{\frac{y}{2\mathcal{R}_\eta\ell}}, \qquad v \simeq \frac{\ell y^2}{2}, \tag{C10}$$

their product becomes

$$xv = \sqrt{\frac{y}{2\mathcal{R}_\eta\ell}}\,\frac{\ell y^2}{2} = \frac{1}{2\sqrt{2}}\sqrt{\frac{\ell}{\mathcal{R}_\eta}}\,y^{5/2}. \tag{C11}$$

The effective temperature therefore takes the asymptotic form

$$\Theta(y) \simeq \sqrt{2\mathcal{R}_\eta\ell}\,F(y) = \sqrt{2\chi}F(y), \tag{C12}$$

where all the remaining frequency dependence is contained in

$$F(y) = \frac{y^4+3y^2+9}{y^{7/2}}. \tag{C13}$$

The prefactor $\sqrt{\ell\mathcal{R}_\eta} = \sqrt{\chi}$ changes the magnitude of the effective temperature but is independent of $y$. Hence, the limiting position of the minimum is determined entirely by the function $F(y)$.

Differentiating,

$$F'(y) = \frac{y^4-9y^2-63}{2y^{9/2}}. \tag{C14}$$

The stationary point therefore satisfies

$$y^4 - 9y^2 - 63 = 0. \tag{C15}$$

Introducing $z = y^2$ gives

$$z^2 - 9z - 63 = 0, \tag{C16}$$

whose positive solution is

$$z = \frac{9+\sqrt{333}}{2}. \tag{C17}$$

The limiting position of the minimum is thus

$$\boxed{y_\infty = \left(\frac{9+\sqrt{333}}{2}\right)^{1/2} \simeq 3.69.} \tag{C18}$$

Since $F(y)$ diverges both as $y \to 0$ and as $y \to \infty$, and the above equation has a single positive root, this stationary point is the global minimum.

Accordingly,

$$\lim_{a\to\infty} y_{\min}(a) = y_\infty. \tag{C19}$$

This saturation applies to the dimensionless position $y_{\min} = a\omega_{\min}/c$, rather than to the dimensional frequency. The latter behaves as

$$\omega_{\min} \simeq \frac{c\,y_\infty}{a}, \tag{C20}$$

and therefore decreases as $a^{-1}$. Physically, the limiting value $y_\infty = O(1)$ reflects the acoustic finite-size condition: the minimum occurs when the longitudinal wavelength is comparable to the particle size, while the finite-$\chi$ viscous and finite-$\ell$ attenuation corrections vanish as the particle radius increases.

We next determine the leading finite-size correction to the position of the acoustic minimum. This correction also identifies the combination of the solvent parameters that controls the approach to the limiting value $y_\infty$.

The corrected dimensionless denominator is

$$\mathcal{D} = 36x^4(1+x)|F_\beta|^2 + \frac{9}{4\mathcal{R}_\eta}|F_\alpha|^2 q^4 \frac{q^2+2v}{v}, \qquad q^2 = u^2+v^2, \tag{C21}$$

whereas the numerator is

$$\begin{aligned} \mathcal{N} &= 9A(x)|F_\beta|^2 \\ &\quad + \frac{9}{2}|F_\alpha|^2\left[D(u,v) + \kappa q^4\left\{2\Lambda(v)(v^2-u^2)+2v+1\right\}\right] + 2\,\mathrm{Re}\,I_{\alpha\beta}, \end{aligned} \tag{C22}$$

with

$$\kappa = \frac{\mu}{2\eta} - \frac{1}{3}, \qquad 1+\kappa = \frac{2}{3} + \frac{\mu}{2\eta} = \frac{1}{2\mathcal{R}_\eta}. \tag{C23}$$

In the large-particle limit, at fixed $y = O(1)$,

$$x = \sqrt{\frac{y}{2\mathcal{R}_\eta \ell}} \gg 1, \qquad u = y + O(\ell^2), \qquad v = \frac{\ell y^2}{2} + O(\ell^3), \tag{C24}$$

and consequently

$$|F_\alpha|^2 = 4x^4 + 12x^3 + O(x^2), \tag{C25a}$$

$$A(x) = 4x^5 + 10x^4 + O(x^3), \tag{C25b}$$

$$|F_\beta|^2 = y^4 + 3y^2 + 9 + O(\ell), \tag{C25c}$$

$$q^2 = y^2 + O(\ell^2). \tag{C25d}$$

The leading contribution to the numerator is therefore the transverse temperature-weighted self-contribution,

$$\mathcal{N}_0(y) = 36x^5\left(y^4+3y^2+9\right). \tag{C26}$$

In contrast, the leading denominator is the part of the longitudinal dissipation that is singular as $v \to 0$,

$$\begin{aligned} \mathcal{D}_0(y) &= \frac{9}{4\mathcal{R}_\eta}(4x^4)y^4\frac{y^2}{v} \\ &= \frac{9}{\mathcal{R}_\eta}\frac{x^4y^6}{v}. \end{aligned} \tag{C27}$$

Thus, the limiting acoustic form is generated by a transverse-dominated numerator and a longitudinal-dominated denominator:

$$\begin{aligned} \Theta_0(y) &= \frac{\mathcal{N}_0(y)}{\mathcal{D}_0(y)} \\ &= 4\mathcal{R}_\eta x v\frac{y^4+3y^2+9}{y^6}. \end{aligned} \tag{C28}$$

Using Eq. (C24),

$$xv = \frac{1}{2\sqrt{2}}\sqrt{\frac{\ell}{\mathcal{R}_\eta}}\,y^{5/2}, \tag{C29}$$

and hence

$$\Theta_0(y) = \sqrt{2\mathcal{R}_\eta \ell}\,F(y) = \sqrt{2\chi}\,F(y), \qquad F(y) = \frac{y^4 + 3y^2 + 9}{y^{7/2}}. \tag{C30}$$

The minimum of $F(y)$ is the limiting position $y_\infty$ derived above.

To calculate the displacement of the minimum, we retain the leading nonanalytic finite-size correction. The terms in $D(u, v)$ that contain the exponential integral can be collected as

$$D_{\log}(u, v) = -2\Lambda(v)(u^2 - v^2)^3. \tag{C31}$$

Similarly, the logarithmic part of the function multiplying $\kappa$ is

$$\begin{aligned} H_{\log}(u, v) &= 2\Lambda(v)q^4(v^2 - u^2) \\ &= -2\Lambda(v)(u^2 + v^2)^2(u^2 - v^2). \end{aligned} \tag{C32}$$

Therefore, in the limit $u \to y$ and $v \to 0$,

$$\begin{aligned} D_{\log} + \kappa H_{\log} &= -2(1 + \kappa)y^6\Lambda(v) + O\big(v^2\Lambda(v)\big) \\ &= -\frac{y^6}{\mathcal{R}_\eta}\Lambda(v) + O\big(v^2\Lambda(v)\big), \end{aligned} \tag{C33}$$

where Eq. (C23) has been used.

The corresponding logarithmic contribution to the numerator is

$$\begin{aligned} \delta\mathcal{N}_{\log} &= \frac{9}{2}|F_\alpha|^2\left[-\frac{y^6}{\mathcal{R}_\eta}\Lambda(v)\right] \\ &= -\frac{18}{\mathcal{R}_\eta}x^4y^6\Lambda(v) + O\big(x^3\Lambda(v)\big). \end{aligned} \tag{C34}$$

Dividing this contribution by the leading denominator in Eq. (C27) gives

$$\begin{aligned} \delta\Theta_{\log}(y) &= \frac{\delta\mathcal{N}_{\log}}{\mathcal{D}_0} \\ &= -2v\Lambda(v). \end{aligned} \tag{C35}$$

For $v \ll 1$,

$$\Lambda(v) = e^{2v}\,\mathrm{Ei}(-2v) = \gamma + \ln(2v) + O(v\ln v), \tag{C36}$$

where $\gamma$ is Euler's constant. Since $2v = \ell y^2 + O(\ell^3)$, Eq. (C35) becomes

$$\delta\Theta_{\log}(y) = -\ell y^2\left[\gamma + \ln(\ell y^2)\right] + O\big(\ell^2\ln\ell\big). \tag{C37}$$

The other corrections arising from finite $x$, the transverse part of the denominator, and the nonlogarithmic longitudinal terms are of order $\ell$ for a fixed solvent. The term in Eq. (C37), which is of order $\ell\ln(1/\ell)$, therefore gives the leading nonanalytic contribution to the shift as $\ell \to 0$.

Combining Eqs. (C30) and (C37), the effective temperature near the limiting minimum can be written as

$$\Theta(y) = \sqrt{2\mathcal{R}_\eta\ell}\,F(y) - \ell y^2\left[\gamma + \ln(\ell y^2)\right] + O(\ell). \tag{C38}$$

Let the position of the finite-size minimum be

$$y_d = y_\infty + \delta y, \qquad |\delta y| \ll 1. \tag{C39}$$

Since $F'(y_\infty) = 0$, differentiation of Eq. (C38) gives

$$0 = \sqrt{2\mathcal{R}_\eta \ell}\, F''(y_\infty)\delta y - 2\ell y_\infty \left[\gamma + \ln(\ell y_\infty^2) + 1\right] + O(\ell). \tag{C40}$$

It follows that

$$\delta y = \frac{\sqrt{2}\, y_\infty}{F''(y_\infty)} \sqrt{\frac{\ell}{\mathcal{R}_\eta}} \left[\gamma + \ln(\ell y_\infty^2) + 1\right]. \tag{C41}$$

For sufficiently small $\ell$, the expression in square brackets is negative, so that $\delta y < 0$ and the finite-size minimum approaches $y_\infty$ from below. Equivalently,

$$y_\infty - y_d = \frac{\sqrt{2}\, y_\infty}{F''(y_\infty)} \sqrt{\frac{\ell}{\mathcal{R}_\eta}} \left[\ln\left(\frac{1}{\ell y_\infty^2}\right) - \gamma - 1\right]. \tag{C42}$$

Using

$$F''(y) = \frac{-y^4 + 45y^2 + 567}{4y^{11/2}}, \tag{C43}$$

and the stationary-point relation $y_\infty^4 = 9y_\infty^2 + 63$, one obtains

$$F''(y_\infty) = \frac{9(y_\infty^2 + 14)}{y_\infty^{11/2}}. \tag{C44}$$

The numerical prefactor in Eq. (C42) is therefore

$$\frac{\sqrt{2}\, y_\infty}{F''(y_\infty)} = \frac{\sqrt{2}\, y_\infty^{13/2}}{9(y_\infty^2 + 14)} \simeq 27.64. \tag{C45}$$

To leading logarithmic accuracy, the finite-size displacement thus has the scaling form

$$y_\infty - y_d \sim \sqrt{\frac{\ell}{\mathcal{R}_\eta}} \ln\left(\frac{1}{\ell}\right). \tag{C46}$$

This identifies the inverse algebraic finite-size variable

$$X_y = \sqrt{\frac{\mathcal{R}_\eta}{\ell}}. \tag{C47}$$

In terms of this variable,

$$y_\infty - y_d \sim \frac{\ln(1/\ell)}{X_y}. \tag{C48}$$

Thus increasing $X_y$, which corresponds to increasing the particle radius, reduces the finite-size correction and drives the minimum towards its asymptotic acoustic position. The logarithmic factor varies slowly compared with the algebraic factor $X_y^{-1}$, explaining the numerical collapse of the dip positions when plotted against $X_y$. A residual logarithmic solvent dependence is nevertheless expected. Indeed, since $\ell = \mathcal{R}_\eta / X_y^2$,

$$y_\infty - y_d \sim \frac{1}{X_y} \left[2 \ln X_y - \ln \mathcal{R}_\eta + \text{constant}\right]. \tag{C49}$$

For comparison, if the bulk-viscous dissipation $\mu|\nabla \cdot \mathbf{v}|^2$ is omitted, the leading longitudinal denominator is instead

$$\mathcal{D}_0^{(\mu\ \text{omitted})} = 12x^4 \frac{y^6}{\nu}. \tag{C50}$$

The corresponding limiting acoustic form is

$$\Theta_0^{(\mu\,\text{omitted})}(y) = \frac{3}{2\sqrt{2}}\sqrt{\frac{\ell}{\mathcal{R}_\eta}}\,F(y). \tag{C51}$$

The logarithmic correction in Eq. (C37) is unchanged at leading order, because both the logarithmic numerator and the singular longitudinal denominator are modified by the same factor. Dividing the correction by the curvature of Eq. (C51) gives

$$y_\infty - y_d \sim \sqrt{\mathcal{R}_\eta \ell}\,\ln\left(\frac{1}{\ell}\right) = \sqrt{\chi}\,\ln\left(\frac{1}{\ell}\right). \tag{C52}$$

The inverse algebraic variable in that incomplete calculation was therefore $\chi^{-1/2}$. Restoring the bulk-viscous dissipation changes the amplitude of the limiting acoustic form from $\sqrt{\ell/\mathcal{R}_\eta}$ to $\sqrt{\mathcal{R}_\eta \ell} = \sqrt{\chi}$, and consequently changes the finite-size scaling variable from $\chi^{-1/2}$ to $\sqrt{\mathcal{R}_\eta/\ell}$.

For a single solvent, $\mathcal{R}_\eta$ is constant and

$$\sqrt{\frac{\mathcal{R}_\eta}{\ell}} = \frac{\mathcal{R}_\eta}{\sqrt{\chi}} \propto \chi^{-1/2}. \tag{C53}$$

Thus both variables have the same particle-size dependence within one solvent. The factor $\mathcal{R}_\eta$ becomes essential when results for solvents with different ratios of shear to longitudinal viscosity, such as water and the Lennard–Jones fluid, are compared.

We next determine the asymptotic width and depth of the acoustic dip. Let $y_-$ denote the local maximum preceding the dip, and let $y_+ > y_d$ be the point on the high-frequency side satisfying

$$\Theta(y_+) = \Theta(y_-). \tag{C54}$$

We define the logarithmic width and the depth by

$$W = \ln\left(\frac{y_+}{y_-}\right), \qquad \Delta\Theta = \Theta(y_-) - \Theta(y_d). \tag{C55}$$

We first consider the left shoulder of the dip. In this region $y \ll 1$, while

$$x = \sqrt{\frac{y}{2\chi}} \gg 1. \tag{C56}$$

The large-$x$ incompressible numerator and denominator have the expansions

$$\mathcal{N}_{\text{inc}} = 324x^5 + 54x^4 + O(x^3), \tag{C57a}$$

$$\mathcal{D}_{\text{inc}} = 324x^5 + 324x^4 + O(x^3). \tag{C57b}$$

Consequently,

$$\Theta_{\text{inc}} = 1 - \frac{5}{6x} + O(x^{-2}). \tag{C58}$$

The corresponding finite-boundary-layer deficiency is

$$1 - \Theta_{\text{inc}} \simeq \frac{5\sqrt{2}}{6}\sqrt{\chi}\,y^{-1/2}. \tag{C59}$$

The singular longitudinal denominator in the same regime is

$$\begin{aligned}\mathcal{D}_L &\simeq \frac{9}{\mathcal{R}_\eta}\frac{x^4 y^6}{v} \\ &= \frac{18}{\chi}x^4 y^4,\end{aligned} \tag{C60}$$

where $v \simeq \ell y^2/2$ and $\chi = \mathcal{R}_\eta \ell$ have been used. The transverse denominator is

$$\mathcal{D}_T \simeq 324x^5. \tag{C61}$$

It follows that

$$\frac{\mathcal{D}_L}{\mathcal{D}_T} \simeq \frac{y^4}{18\chi x} = \frac{\sqrt{2}}{18}\frac{y^{7/2}}{\sqrt{\chi}}. \tag{C62}$$

Thus, to leading order, the deficiency of the effective temperature on the low-frequency side is

$$1 - \Theta(y) \simeq \frac{5\sqrt{2}}{6}\sqrt{\chi}\, y^{-1/2} + \frac{\sqrt{2}}{18}\frac{y^{7/2}}{\sqrt{\chi}}. \tag{C63}$$

The first term decreases with frequency, whereas the second increases. The maximum $y_-$ is obtained by differentiating Eq. (C63):

$$-\frac{5\sqrt{2}}{12}\sqrt{\chi}\, y_-^{-3/2} + \frac{7\sqrt{2}}{36}\frac{y_-^{5/2}}{\sqrt{\chi}} = 0. \tag{C64}$$

Therefore,

$$y_-^4 = \frac{15}{7}\chi, \qquad y_- = \left(\frac{15}{7}\right)^{1/4}\chi^{1/4}. \tag{C65}$$

For a fixed solvent this is equivalent to $y_- \propto \ell^{1/4}$.

At the stationary point, the first contribution in Eq. (C63) is seven times the second. Hence

$$\Theta(y_-) = 1 - C_-\chi^{3/8} + o(\chi^{3/8}), \tag{C66}$$

where

$$C_- = \frac{8}{7}\frac{5\sqrt{2}}{6}\left(\frac{15}{7}\right)^{-1/8} \simeq 1.22448. \tag{C67}$$

#### 1. The high frequency and the intermerdiate frequency on the right side of the dip for $y > y_d$

We next examine the high-frequency side of the dip. It is important to distinguish the ultimate high-frequency limit from the intermediate acoustic regime that contains the right equal-height point $y_+$. Although both regimes lie on the high-frequency side of the minimum, they correspond to different asymptotic forms of the longitudinal wave number.

The dimensionless longitudinal wave number is

$$\widehat{\beta} \equiv a\beta = u + iv = \frac{y}{\sqrt{1 - i\ell y}}, \qquad y = \frac{a\omega}{c}. \tag{C21}$$

Writing

$$r_\ell = \sqrt{1 + \ell^2 y^2}, \tag{C22}$$

its real and imaginary parts may be expressed exactly as

$$u^2 = \frac{y^2(r_\ell + 1)}{2(1 + \ell^2 y^2)}, \tag{C23}$$

$$v^2 = \frac{y^2(r_\ell - 1)}{2(1 + \ell^2 y^2)}. \tag{C24}$$

The wavelength of the longitudinal mode and its amplitude attenuation length, relative to the particle radius, are respectively

$$\frac{\lambda_\beta}{a} = \frac{2\pi}{u}, \qquad \frac{L_\beta}{a} = \frac{1}{v}. \tag{C25}$$

Since the dissipation density contains the squared velocity amplitude, its radial decay length is instead $a/(2v)$.

In the ultimate high-frequency limit,

$$\ell y \gg 1, \tag{C26}$$

the longitudinal wave number becomes

$$u \simeq v \simeq \sqrt{\frac{y}{2\ell}}. \tag{C27}$$

Thus, $u$ and $v$ increase as the square root of the frequency, rather than linearly with the frequency. The transverse wave number is

$$x = \sqrt{\frac{y}{2\chi}}, \tag{C28}$$

and, using $\mathcal{R}_\eta = \chi/\ell$, one finds

$$v \simeq \sqrt{\mathcal{R}_\eta}\, x. \tag{C29}$$

Both the longitudinal and transverse dissipation fields are therefore confined to thin layers adjacent to the particle surface. Since the temperature weighting $a/r$ differs only weakly from unity within these layers, the effective temperature approaches the particle surface temperature. The corrected high-frequency expansion is

$$\Theta(y) = 1 - \frac{2s^2+1}{2s(2s+1)}\frac{1}{x} + O(x^{-2}), \qquad s = \sqrt{\mathcal{R}_\eta}. \tag{C30}$$

Equivalently,

$$1 - \Theta(y) = \frac{2s^2+1}{2(2s+1)}\frac{1}{v} + O(v^{-2}). \tag{C31}$$

The $x^{-1}$ behavior therefore describes the ultimate $y \to \infty$ limit.

Upon decreasing the frequency, the system crosses into the weakly attenuated acoustic regime

$$\ell y \ll 1. \tag{C32}$$

Expanding the longitudinal wave number gives

$$u = y + O(\ell^2 y^3), \qquad v = \frac{\ell y^2}{2} + O(\ell^3 y^4). \tag{C33}$$

The wavelength and attenuation length are then

$$\frac{\lambda_\beta}{a} \simeq \frac{2\pi}{y}, \qquad \frac{L_\beta}{a} \simeq \frac{2}{\ell y^2}. \tag{C34}$$

Both lengths increase as the frequency is reduced, but the attenuation length increases faster, as $y^{-2}$, than the wavelength, which increases as $y^{-1}$.

There exists an intermediate range on the right side of the dip in which

$$\ell y \ll 1, \qquad v = \frac{\ell y^2}{2} \gg 1. \tag{C35}$$

Equivalently,

$$\ell^{-1/2} \ll y \ll \ell^{-1}. \tag{C36}$$

In this regime the longitudinal mode is weakly attenuated per wavelength, since $v/u \simeq \ell y/2 \ll 1$, while its dissipation is still localized within a layer much thinner than the particle radius, since $v \gg 1$.

To determine the effective temperature in this region, write

$$R = \frac{r}{a} = 1 + z. \tag{C37}$$

The dominant longitudinal contribution to the unweighted dissipation has the form

$$\mathcal{D}_L \propto \int_0^\infty e^{-2vz}\Psi(1+z)\,dz, \tag{C38}$$

where $\Psi(R)$ contains the algebraic factors arising from the velocity gradients. The corresponding temperature-weighted numerator contains the additional factor $1/R$:

$$\mathcal{N}_L \propto \int_0^\infty \frac{e^{-2vz}\Psi(1+z)}{1+z}\,dz. \tag{C39}$$

For $v \gg 1$, the integrals are dominated by $z = O(v^{-1})$. We may therefore expand

$$\Psi(1+z) = \Psi_0 + \Psi_1 z + O(z^2), \tag{C40}$$

and

$$\frac{1}{1+z} = 1 - z + O(z^2). \tag{C41}$$

Using

$$\int_0^\infty e^{-2vz}dz = \frac{1}{2v}, \qquad \int_0^\infty ze^{-2vz}dz = \frac{1}{(2v)^2}, \tag{C42}$$

we obtain

$$\mathcal{D}_L = \frac{\Psi_0}{2v} + \frac{\Psi_1}{(2v)^2} + O(v^{-3}), \tag{C43}$$

$$\mathcal{N}_L = \frac{\Psi_0}{2v} + \frac{\Psi_1 - \Psi_0}{(2v)^2} + O(v^{-3}). \tag{C44}$$

Consequently,

$$\frac{\mathcal{N}_L}{\mathcal{D}_L} = 1 - \frac{1}{2v} + O(v^{-2}). \tag{C45}$$

The correction has a simple spatial interpretation. The longitudinal dissipation is localized at a mean dimensionless distance

$$\langle R - 1\rangle \simeq \frac{1}{2v} \tag{C46}$$

from the particle surface. Since the temperature weighting is $1/R \simeq 1 - (R-1)$, the effective-temperature deficit is equal, to leading order, to this mean displacement.

In the intermediate acoustic regime, the longitudinal contribution also dominates the transverse contribution. Indeed, using $u \simeq y$, $v \simeq \ell y^2/2$, and $x^2 = y/(2\chi)$, one finds parametrically

$$\frac{\mathcal{D}_T}{\mathcal{D}_L} \sim \sqrt{\ell y}, \tag{C47}$$

which tends to zero when $\ell y \to 0$. Hence the full effective temperature has the asymptotic form

$$\Theta(y) = 1 - \frac{1}{2v} + O\left(v^{-2}\right) + O\left(\sqrt{\ell y}\right). \tag{C48}$$

Substituting $2v \simeq \ell y^2$ gives

$$1 - \Theta(y) \simeq \frac{1}{\ell y^2}. \tag{C49}$$

This expression therefore applies in the intermediate window

$$\ell^{-1/2} \ll y \ll \ell^{-1}, \tag{C50}$$

and should not be identified with the ultimate $y \to \infty$ result $1 - \Theta \propto x^{-1}$.

The two asymptotic forms are mutually consistent. Their crossover occurs when

$$\ell y = O(1), \qquad y = O(\ell^{-1}). \tag{C51}$$

At this frequency, the intermediate result gives

$$\frac{1}{\ell y^2} = O(\ell). \tag{C52}$$

At the same point,

$$x = \sqrt{\frac{1}{2\chi\ell}} = \frac{1}{\sqrt{2\mathcal{R}_\eta}\,\ell}, \tag{C53}$$

and hence

$$\frac{1}{x} = O(\ell). \tag{C54}$$

Thus both expressions have the same order of magnitude at the crossover, although their numerical coefficients are different because neither limiting expansion is uniformly accurate when $\ell y = O(1)$.

Starting from the ultimate high-frequency side and decreasing the frequency, the physical sequence is therefore as follows. For $\ell y \gg 1$, both the transverse and longitudinal dissipation are confined close to the hot surface and $1 - \Theta \propto x^{-1}$. On decreasing $y$ through $\ell y = O(1)$, the longitudinal mode enters the weakly attenuated acoustic regime. Its attenuation length increases and the associated dissipation extends farther into the colder solvent. In the intermediate range $\ell^{-1/2} \ll y \ll \ell^{-1}$, this outward displacement produces

$$1 - \Theta \simeq \frac{1}{\ell y^2}, \tag{C55}$$

so that $\Theta$ continues to decrease as the frequency is lowered.

When $y$ is reduced to $O(1)$, the longitudinal wavelength becomes comparable to the particle size,

$$\frac{\lambda_\beta}{a} \simeq \frac{2\pi}{y} = O(1). \tag{C56}$$

At the same time, $\nu \simeq \ell y^2/2 \ll 1$, so the acoustic field extends far into the colder solvent. The effective temperature is then determined by the ratio of the transverse-dominated temperature-weighted numerator to the attenuation-singular longitudinal denominator,

$$\Theta_0(y) = \sqrt{2\chi}\,\frac{y^4 + 3y^2 + 9}{y^{7/2}}. \tag{C57}$$

The minimum of this acoustic form occurs at

$$y_\infty = \sqrt{\frac{9 + \sqrt{333}}{2}} \simeq 3.69. \tag{C58}$$

The minimum is therefore not produced simply by the attenuation length becoming largest. Rather, it occurs where the longitudinal field is sufficiently extended to sample the colder solvent while the particle still couples strongly to the acoustic mode.

On decreasing the frequency below the minimum, the acoustic wavelength becomes much larger than the particle size. The particle then becomes acoustically small and the fraction of the total dissipation carried by the longitudinal mode decreases. In the low-frequency-side regime $y \ll 1$, but $x \gg 1$, the ratio of the longitudinal and transverse dissipation behaves as

$$\frac{\mathcal{D}_L}{\mathcal{D}_T} \simeq \frac{\sqrt{2}}{18}\frac{y^{7/2}}{\sqrt{\chi}}. \tag{C59}$$

It therefore vanishes rapidly as $y$ is decreased. Although the longitudinal attenuation length continues to increase, the longitudinal mode carries an increasingly small fraction of the total dissipation. The transverse viscous contribution consequently takes over and the effective temperature rises from the acoustic minimum toward its low-frequency value.

The minimum thus reflects a competition between two effects. On the high-frequency side, the acoustic mode carries substantial dissipation but is localized close to the hot surface. On the low-frequency side, the acoustic field is spatially extended but carries only a small fraction of the total dissipation. The largest relative weighting of the colder solvent is obtained between these limits, when $y = O(1)$.

### 2. Calculation of $y_+$

At the equal-height point $y_+$, the longitudinal attenuation parameter satisfies $v_+ \gg 1$. Writing $R = 1 + z$, the longitudinal denominator and temperature-weighted numerator have the generic forms

$$\mathcal{D}_L \propto \int_0^\infty e^{-2vz}\Psi(1+z)\,dz, \tag{C68a}$$

$$\mathcal{N}_L \propto \int_0^\infty \frac{e^{-2vz}\Psi(1+z)}{1+z}\,dz. \tag{C68b}$$

Expanding $\Psi(1+z) = \Psi_0 + \Psi_1 z + \cdots$ and $(1+z)^{-1} = 1 - z + \cdots$, we obtain

$$\mathcal{D}_L = \frac{\Psi_0}{2v} + \frac{\Psi_1}{(2v)^2} + O(v^{-3}), \tag{C69a}$$

$$\mathcal{N}_L = \frac{\Psi_0}{2v} + \frac{\Psi_1 - \Psi_0}{(2v)^2} + O(v^{-3}). \tag{C69b}$$

Their ratio is therefore

$$\Theta(y) = 1 - \frac{1}{2v} + O(v^{-2}) = 1 - \frac{1}{\ell y^2} + O(v^{-2}). \tag{C70}$$

The equal-height condition $\Theta(y_+) = \Theta(y_-)$ now gives

$$\frac{1}{\ell y_+^2} = C_- \chi^{3/8}. \tag{C71}$$

Consequently,

$$\begin{aligned} y_+ &= C_-^{-1/2} \ell^{-1/2} \chi^{-3/16} \\ &= C_-^{-1/2} \mathcal{R}_\eta^{-3/16} \ell^{-11/16}. \end{aligned} \tag{C72}$$

Thus, for a fixed solvent, $y_+ \propto \ell^{-11/16}$.

Combining Eqs. (C65) and (C72), the logarithmic width is

$$\begin{aligned} W &= \frac{1}{2}\ln\left(\frac{1}{\ell}\right) + \frac{7}{16}\ln\left(\frac{1}{\chi}\right) + C_W \\ &= \frac{15}{16}\ln\left(\frac{1}{\ell}\right) - \frac{7}{16}\ln \mathcal{R}_\eta + C_W, \end{aligned} \tag{C73}$$

where

$$C_W = -\frac{1}{2}\ln C_- - \frac{1}{4}\ln\left(\frac{15}{7}\right). \tag{C74}$$

For a fixed solvent, the asymptotic slope of $W$ against $\ln(1/\ell)$ is therefore $15/16$. The term $-(7/16)\ln \mathcal{R}_\eta$ represents a solvent-dependent offset.

Finally, the minimum value follows from the limiting acoustic form,

$$\Theta(y_d) = C_d \sqrt{\chi} + o(\sqrt{\chi}), \tag{C75}$$

where

$$C_d = \sqrt{2} F(y_\infty) \simeq 3.44704. \tag{C76}$$

Combining Eqs. (C66) and (C75), the depth becomes

$$\Delta\Theta = 1 - C_- \chi^{3/8} - C_d \chi^{1/2} + \cdots. \tag{C77}$$

Thus the strict leading correction to the limiting depth is

$$1 - \Delta\Theta \sim \chi^{3/8}. \tag{C78}$$

Introducing $X_\Delta = \chi^{-1/2}$, Eq. (C77) may be written as

$$\Delta\Theta = 1 - C_- X_\Delta^{-3/4} - C_d X_\Delta^{-1} + \cdots. \tag{C79}$$

This identifies $\chi^{-1/2}$ as the natural particle-size variable governing the depth of the acoustic dip.



\end{document}